\documentclass[12pt]{article}

\usepackage[utf8]{inputenc}   
\usepackage{csquotes}

\usepackage[utf8]{inputenc}
\usepackage[english]{babel}
\usepackage{authblk}

\usepackage[a4paper, margin=1in]{geometry}

\DeclareUnicodeCharacter{03C0}{\ensuremath{\pi}}
\DeclareUnicodeCharacter{2009}{\,}

\usepackage[
    backend=biber,
    style=numeric-comp, 
    sorting=none
]{biblatex}
\AtEveryBibitem{\clearfield{note}}
\usepackage{amsmath, amssymb}
\usepackage{graphicx}

\usepackage[colorlinks=true, linkcolor=blue, citecolor=blue, urlcolor=blue]{hyperref}
\title{Elucidating Norrish Type-I reactive pathways by ultrafast X-ray absorption spectroscopy}

\author[1,2]{Martin Gra{\ss}l} 
\author[1,3]{Pablo Unzueta} 
\author[1,3]{Andreas E. Hillers-Bendtsen} 
\author[1,4]{Yusong Liu}
\author[1,3,5,6]{Diptarka Hait}
\author[1,7,8]{Alice E. Green}
\author[4]{Xinxin Cheng}
\author[1, 9]{Felix Allum}
\author[1,4]{Taran Driver}
\author[4,10]{Ruaridh Forbes}
\author[4]{James. M. Glownia}
\author[1]{Erik Isele}
\author[1,4]{Kirk A. Larsen}
\author[4]{Xiang Li}
\author[4]{Ming-Fu Lin}
\author[4]{Razib Obaid}
\author[1,4]{Adam Summers}
\author[1]{Emily Thierstein}
\author[1]{Jun Wang}
\author[1,4]{James P. Cryan}
\author[1,2,4]{Matthias F. Kling}
\author[1,3]{Todd J. Martinez}
\author[1,4]{Thomas J. A. Wolf}

\affil[1]{Stanford PULSE Institute, SLAC National Accelerator Laboratory, 2575 Sand Hill Road, Menlo Park, CA 94025, USA.}
\affil[2]{Department of Physics, Ludwig-Maximilians-Universität München, Am Coulombwall 1, D-85748 Garching, Germany.}
\affil[3]{Department of Chemistry, Stanford University, 333 Campus Drive, Stanford, CA 94305, USA}
\affil[4]{Linac Coherent Light Source, SLAC National Accelerator Laboratory, 2575 Sand Hill Road, Menlo Park, CA 94025, USA.}
\affil[5]{Department of Chemistry, Columbia University, New York, New York 10027, United States.}
\affil[6]{Initiative for Computational Catalysis, Flatiron Institute, New York, New York 10010, United States.}
\affil[7]{European XFEL, Holzkoppel 4, 22869 Schenefeld, Germany.}
\affil[8]{EaStCHEM School of Chemistry, University of Edinburgh, Edinburgh EH9 3FJ, U.K.}
\affil[9]{Deutsches Elektronen-Synchrotron DESY, Notkestr. 85, 22607 Hamburg, Germany.}
\affil[10]{Department of Chemistry, University of California, Davis, California 95616, USA.}

\begin{document}

\maketitle

Norrish type I reactions selectively cleave carbon-carbon bonds directly adjacent to carbonyl groups. Despite their broad use in combination with aromatic carbonyls for additive manufacturing and dental UV curing applications, the nature of the photochemically active state and its population mechanism remain insufficiently understood. Detailed mechanistic insight requires mapping of the photoexcited population flow involving internal conversion and intersystem crossing. We present a time-domain study of gas phase acetophenone as a prototypical aromatic carbonyl combining soft X-ray time-resolved near-edge X-ray absorption fine structure (TR-NEXAFS) spectroscopy at the oxygen K-edge with ab initio multiple spawning (AIMS) simulations. Exploiting the specific sensitivity of TR-NEXAFS spectroscopy to states with $n\pi^*$ character, we observe population transfer from the initially excited $^1\pi\pi^*$ state to the $^1n\pi^*$ state with a time constant of $(0.13 \pm 0.02)$ ps after an initial induction period of $(0.12 \pm 0.02)$ ps  without population transfer, in quantitative agreement with the AIMS simulations. The population in the $^1n\pi^*$ state subsequently decays via intersystem crossing, likely mediated by a $^3\pi\pi^*$ state,  within $(3.17 \pm 0.66)$ ps to a long-lived $^3n\pi^*$ state, which is presumed to be active towards Norrish type I chemistry.

\section{Introduction}
The efficient conversion of light into chemical and mechanical energy is pivotal to many chemical and biological reactions. It often involves radiationless de-excitation processes taking place on ultrafast timescales and is mediated by the coupling of electronic and nuclear dynamics, which cannot be described within the Born-Oppenheimer approximation. Such nonadiabatic dynamics in the vicinity of conical intersections (CIs) still challenge experimental and quantum chemical investigation approaches. In many cases, the challenge in predicting and understanding ultrafast photochemical reaction mechanisms is further augmented by the presence of intersystem crossing (ISC) pathways providing the photoexcited molecules access to different spin state manifolds. An example is the class of aromatic carbonyls, which exhibit rich photochemistry including Norrish type I and type II, hydrogen atom transfer, and Paterno-Buechi, reactions. \cite{padwa_organic_2017, dinda_essentials_2017} Specifically, their Norrish type reactivity is widely used in synthetic applications \cite{karkas_photochemical_2016} and photopolymerization\cite{muller_recent_2022} with applications in additive manufacturing\cite{bao_recent_2022} and dentistry.\cite{rueggeberg_state---art_2011}

It is well-established that the Norrish-type reactions of aromatic carbonyls, following $\pi\pi^*$ excitation, take place in their triplet manifolds enabled by close-to-unity ISC quantum yields.\cite{gilbert_essentials_1991} The high efficiency of spin-forbidden ISC is enabled by the presence of excited singlet and triplet states with $\pi\pi^*$ and $n\pi^*$ character, which are close in energy and exhibit strong spin-orbit coupling according to the El Sayed selection rule.\cite{el_sayed_spin-orbit_1963} However, it is unclear if the triplet manifold is accessed directly via $^1\pi\pi^*/^3n\pi^*$ ISC or indirectly via $^1\pi\pi^*/^1n\pi^*$ internal conversion (IC) followed by $^1n\pi^*/^3\pi\pi^*$ ISC (see Fig.~\ref{fig:ExpSetup} c). Additionally, the electronic character of the state in which the Norrish type I dissociation takes place is unclear. The nature of the access to the triplet manifold and of the populated states has important consequences for competing processes, such as IC to the electronic ground state, and the potential of their control through modifications of the reactant.

A promising approach to gaining a detailed understanding of ultrafast photochemical reaction mechanisms is the investigation of comparably small, organic, and isolated model systems in the gas phase with specifically targeted experimental observables.\cite{wolf_probing_2017, facciala_unraveling_2025, mayer_following_2022, stankus_ultrafast_2019, wolf_photochemical_2019,blanchet_discerning_1999, attar_femtosecond_2017,pertot_time-resolved_2017, li_imaging_2025} This approach also allows for detailed comparisons to quantum chemical simulations of these time-dependent experimental observables.\cite{bennett_prediction_2024, eng_photochemistry_2024, hait_prediction_2024, hutton_using_2024, jaiswal_ultrafast_2024, janos_predicting_2024, lawrence_mash_2024, makhov_ultrafast_2024, martin_santa_daria_photofragmentation_2024, miao_casscfmrci_2024, miller_ultrafast_2024, mukherjee_prediction_2024, peng_photodissociation_2024, suchan_prediction_2024, vindel-zandbergen_non-adiabatic_2024, wang_imaging_2025, green_imaging_2025, chakraborty_time-resolved_2024} Such comparisons have the promise to lead to a predictive understanding of ultrafast photochemistry in the future.

Here, we investigate these pathways in a prototypical aromatic carbonyl, acetophenone (AP), excited at 267~nm. The structure of AP is shown in Fig.~\ref{fig:ExpSetup}. Its absorption band associated with the $\pi\pi^*$ transition has its maximum at 275~nm and is well-separated from a higher-lying $\pi\pi^*$ transition centered at 238~nm.\cite{berger_photochemical_1975} Additionally, it exhibits a very weak band at 320~nm, which is associated with its $^1n\pi^*$ state. Photoexcitation of the $^1\pi\pi^*$ state of AP leads to ISC and to the formation of phenylcarbonyl and methyl radicals via a Norrish type I reaction with unity quantum yield and a time constant of 2~ns.\cite{berger_photochemical_1975, zhao_laser_1997}. An alternative Norrish type I reaction channel, yielding phenyl and acetaldehyde radicals, has not been observed. Intersystem crossing was inferred to take place within 260 fs after $^1\pi\pi^*$ excitation from the bandwidth of the corresponding line in photoexcitation spectroscopy of jet-cooled AP.\cite{warren_s2_1986} The first time-domain investigation of gas phase AP was performed using time-resolved photoelectron spectroscopy.\cite{lee_substituent_2002} A lifetime of 140~fs for the $^1\pi\pi^*$ state was obtained. Additionally, the structural dynamics of AP were investigated by ultrafast electron diffraction (UED).\cite{feenstra_excited_2005, park_ultrafast_2006} The UED study finds a channel bifurcation for the photoinduced process. The $^3\pi\pi^*$ state is populated within 50 ps. Additionally, the study finds a Norrish type I reaction channel directly from the $^1n\pi^*$ singlet state within 100 ps. 

These previous results leave unanswered questions for Norrish type I reaction dynamics of AP and of aromatic carbonyls in general, specifically about (1) the role of the $^1n\pi^*$ state, (2) its lifetime, and (3) the character of the reactive triplet state. We address these gaps with a combined experimental and simulation study in the gas phase using soft X-ray time-resolved near-edge X-ray absorption fine structure (NEXAFS) spectroscopy at the oxygen edge and \textit{ab initio} multiple spawning (AIMS) simulations.\cite{ben-nun_ab_2000} 

It was previously demonstrated that time-resolved NEXAFS spectroscopy can provide selective sensitivity to population in excited states of $n \pi^*$ character.\cite{wolf_probing_2017, kjonstad_photoinduced_2024} This selectivity arises from the strong dependence of NEXAFS cross-sections on the spatial overlap of core and valence orbital densities. Hence, the strongly localized nature of the lone pair electron hole in the $n\pi^*$ state at the heteroatom site (see Fig.~\ref{fig:ExpSetup} b) yields substantially larger cross-sections for resonances with the heteroatom core orbitals compared to, for example, strongly delocalized $\pi$ electron holes in states of $\pi\pi^*$ character. 

The AIMS simulations are performed using the hole-hole Tamm-Dancoff approximated (\textit{hh}-TDA) density functional theory electronic structure method, which has been shown to be an accurate and efficient technique for molecules with low-lying $n\pi^*$ and $\pi\pi^*$ excited states, and also amenable to simulation of transient NEXAFS spectra \cite{bannwarth_holehole_2020,yu2020ab,hohenstein_predictions_2021}.

\section{Experimental}\label{methods}
\begin{figure}[!ht]
    \centering
    \includegraphics[width=\columnwidth]{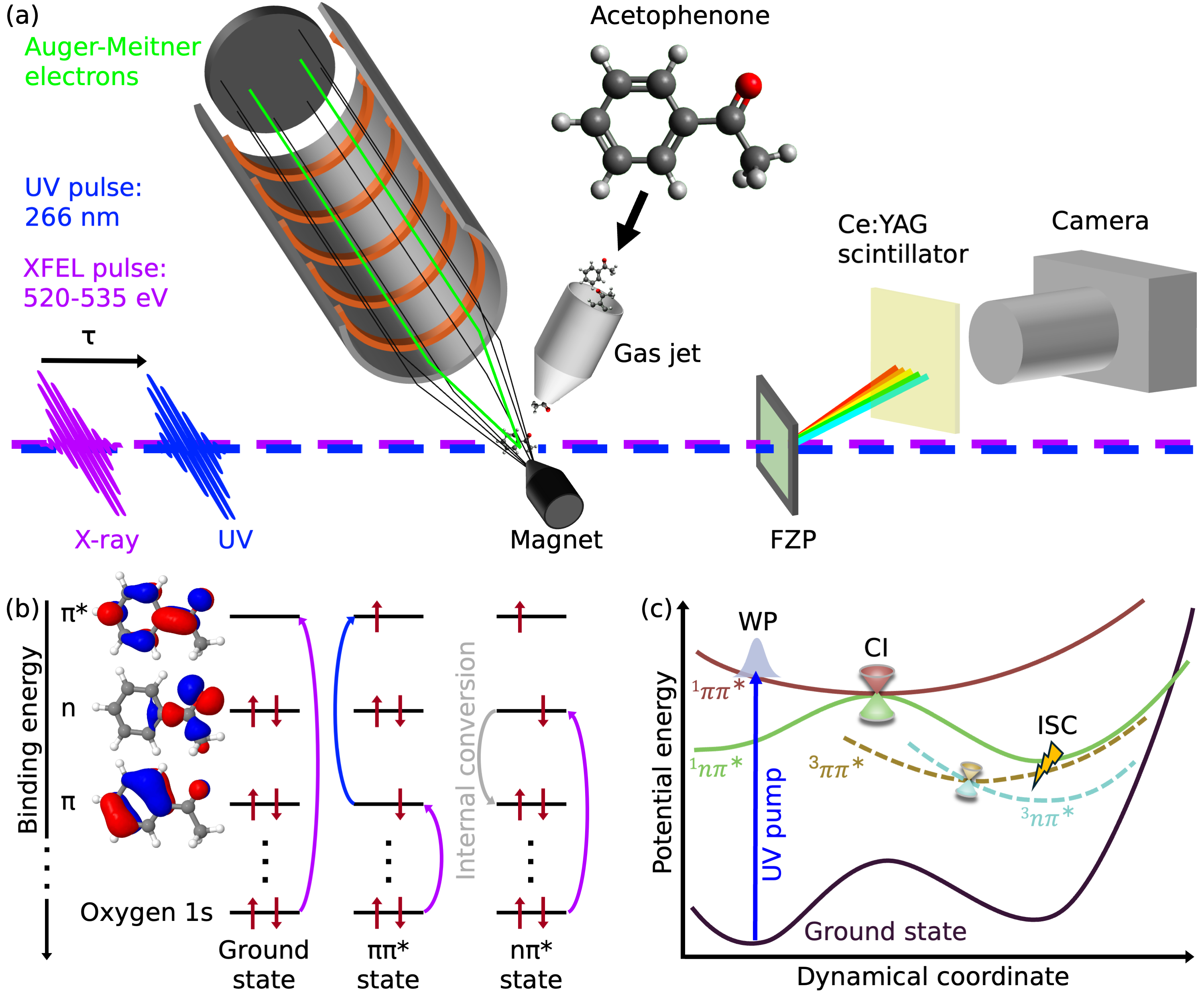}
    \caption{Schematic of the experimental setup. a) A 266~nm UV laser pulse (blue) excites the acetophenone sample with a variable delay $\tau$ relative to the X-ray pulse (purple), which has a tunable photon energy between 520 and 535~eV. The sample is injected into the interaction region using a gas jet, and resonant Auger–Meitner electrons produced by the X-ray-sample interaction are collected by a magnetic bottle spectrometer. The X-ray pulse spectrum is measured by dispersing the beam with a Fresnel zone plate (FZP) onto a Ce:YAG scintillator. b) Electron configurations (right) and isosurface representation (left) of the natural transition orbitals defining the the $^1n\pi^{*}$ and $^1\pi\pi^{*}$ electronic character of the relevant excited states. In the ground state, all molecular orbitals are filled up to the orbital with $n$ character. The lowest unoccupied molecular orbital has $\pi^{*}$ character. The UV laser excites the molecule into the lowest $^1\pi\pi^{*}$ state by promoting one electron from an occupied molecular orbital with $\pi$ character into the $\pi^*$ orbital (blue arrow). Through internal conversion the molecule can transition into a $^1n\pi^{*}$ state, where an electron from the n orbital fills the hole in the $\pi$ orbital. The X-rays can resonantly excite oxygen 1s electrons into valence orbital vacancies (purple arrow). c) Sketch of the proposed reaction pathway leading from the the $^1\pi\pi^{*}$ state through a conical intersection (CI) between to the $^1n\pi^{*}$ state, followed by intersystem crossing (ISC) to the $^3\pi\pi^{*}$ state and subsequent internal conversion to the $^3n\pi^{*}$ state.}
    \label{fig:ExpSetup}
\end{figure}
Time-resolved NEXAFS spectroscopy is performed at the Linac Coherent Light Source (LCLS) free electron laser (FEL) facility, SLAC National Accelerator Laboratory, using the time-resolved atomic, molecular and optical science (TMO) instrument \cite{walter_time-resolved_2022}. A schematic representation of the experimental setup is shown in Figure \ref{fig:ExpSetup} a). The sample is purchased from Sigma Aldrich and evaporated using a bubbler seeded with helium through a gas needle directly into the overlap region of the optical and X-ray laser in an ultra high vacuum chamber. The sample is excited by 266~nm laser pulses with a duration of 45~fs full width at half maximum (FWHM), a focus diameter of 212~$\mu$m FWHM, and a pulse energy of 18.2~$\mu$J. Soft X-ray pulses with 10~fs duration and an average bandwidth of 2.6~eV FWHM are focused to a diameter of 12~um FWHM using a pair of Kirkpatrick-Baez mirrors and used to probe the sample in the pre-edge region of the oxygen K-edge between 520 and 535~eV. Their central photon energy is tuned by varying the gaps of the undulators. We take advantage of the strongly fluctuating spectral distribution of self-amplified spontaneous emission (SASE) pulses created by LCLS by applying the spectral domain ghost imaging technique to improve the energy resolution to $\approx$0.1~eV.\cite{driver_attosecond_2024} Thereby the absorbance of the sample is correlated with the photon spectrum of the X-rays on a pulse-by-pulse basis \cite{li_time-resolved_2021, li_two-dimensional_2021}. The absorbance is measured by detecting the Auger-Meitner electron yield, which is proportional to the photon energy-dependent absorption cross-section of the sample, with a magnetic bottle spectrometer \cite{borne_design_2024}. The spectrum of the FEL is obtained on a shot-by-shot basis using an in-line Fresnel zone plate (FZP) spectrometer downstream from the magnetic bottle spectrometer \cite{larsen_compact_2023}. The relative time delay between UV and X-ray pulses is measured on a shot-by-shot basis with an arrival time monitor (ATM) \cite{droste_high-sensitivity_2020}.  To ensure that the experiment takes place in the linear absorption regime, the intensity of the transient signal at 527~eV is measured for a range of UV pump intensities (see the Supplementary Information (SI), section 3, for details). X-ray pulses are delayed with respect to UV laser pulses between -2 ps and 20~ps. The transient NEXAFS spectra are obtained by evaluating the difference signal, where the average of the first four time steps preceding $t_0$ is used as a background reference. The dataset of individual ATM-tagged shots is sorted into delay bins of $\approx$ 60 fs for the region between -2 ps and 3~ps by varying the exact bin-width to maintain a comparable number of shots for the ghost imaging evaluation. Additionally, transient spectra are obtained in separate measurements at delays of 5, 10, and 20 ps. The NEXAFS spectra are calibrated using the literature spectra of benzaldehyde and CO$_2$ \cite{wight_k-shell_1974, hitchcock_inner-shell_1992} (see the SI,  section 2).

The nonadiabatic dynamics within the singlet state manifold were simulated using the AIMS method\cite{ben1998nonadiabatic,ben-nun_ab_2000} in combination with electronic structure calculations employing the fomo-hh-TDA-BHandHLYP/def2-SVP methodology\cite{bannwarth_holehole_2020,yu2020ab,weigend2005balanced,becke1988density,lee1988development,becke1993new} with constant smearing of the electrons across the active orbitals as implemented in TeraChem\cite{seritan2020terachem,seritan2021terachem}. 50 initial conditions were sampled from a 0~K Wigner distribution\cite{wigner1932quantum} and propagated for 45000 a.u. (corresponding to 1.089~ps) with a timestep of 0.5~fs except for regions with high nonadiabatic coupling, where the stepsize was reduced to 0.25~fs. The same electronic structure level was also employed to generate time-resolved NEXAFS spectra from the AIMS simulations, where we only compute the spectrum for the first 500~fs, since the electronic state populations barely change beyond this point. The static NEXAFS spectra of the ground state, the $^1n\pi^*$ state, and the $^3n\pi^*$ state were compared using the OO-DFT/SCAN level of theory\cite{sun2015strongly,hait2021orbital} with scalar relativistic effects included using the spin-free exact two component model\cite{cunha2022relativistic} in combination with the aug-pcX-2 basis\cite{ambroise2018probing} on Oxygen and the aug-pcseg-1 basis\cite{jensen2014unifying} on all other atoms as implemented in QChem\cite{epifanovsky2021software}, using the respective S$_0$, S$_1$, and T$_1$ BHandHLYP/def2-SVP optimized geometries. To facilitate comparison with the experimental data, the simulated spectrum is convolved with a 50~fs Gaussian function to account for the experimental instrument response function. Furthermore, an offset of 4.0~eV is applied to match the maxima of simulated (527.0~eV) and experimental (531.0~eV) $\pi^*$ resonance signatures in the ground state spectra. Additional details on the simulation methodology can be found in the SI section 5. 

\section{Results and discussion}

\begin{figure}[!ht]
    \centering
    \includegraphics[width=0.5\textwidth]{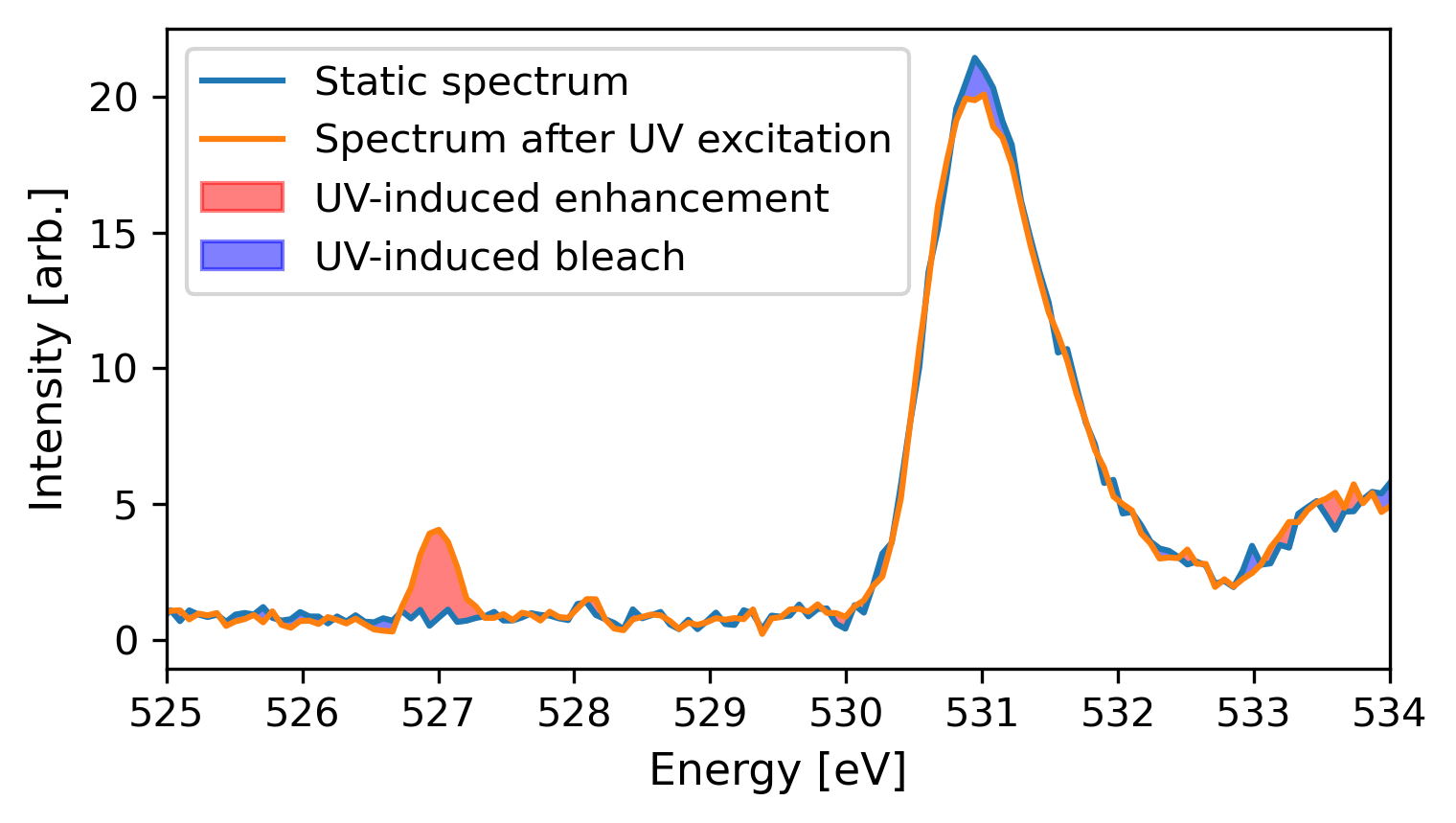}
    \caption{Acetophenone absorption spectrum before (blue) and 0.5 ps after UV excitation (orange). UV excitation induces a new peak at 527~eV and a depletion to the main peak of the ground state spectrum at 531~eV, shown by the red and blue areas, respectively.}
    \label{fig:signalvsbkg}
\end{figure}

\subsection{Ground state and excited state NEXAFS spectra} 
The blue line in Figure \ref{fig:signalvsbkg} shows the experimental ground state NEXAFS spectrum of AP. The energy resolution of the spectrum ($\approx$ 0.1~eV) is significantly beyond the bandwidth of the FEL pulses ($\approx$ 2.6~eV) and is obtained using spectral domain ghost imaging (see the section \ref{methods} and additional details in the SI section 1) \cite{li_two-dimensional_2021}. The main peak of the spectrum at 531.0~eV corresponds to the $\pi^*$ resonance of AP, a resonant transition of an oxygen 1s electron into the lowest unoccupied molecular orbital (LUMO), which has carbonyl $\pi^*$ character (see Figure \ref{fig:ExpSetup}b)). The overall shape of the measured spectrum closely resembles the spectrum of the structurally similar benzaldehyde molecule reported in the literature \cite{hitchcock_inner-shell_1992}. See the SI section 2 for a complete characterization of the spectrum.  

The orange line in Figure \ref{fig:signalvsbkg} shows the NEXAFS spectrum taken 0.5~ps after photoexcitation at 266~nm. In comparison to the static spectrum, the excitation leads to the appearance of an additional red-shifted peak at 527.0~eV and to a depletion of the ground state absorption peak at 531.0~eV, illustrated by the red and blue areas, respectively. The observed UV-induced features can be explained by a change in electron configuration in the excited sample. The UV laser excites a valence electron into a previously unoccupied molecular orbital. The single electron transition results in an electron hole in a formerly occupied molecular orbital, which in turn enables a new, red-shifted resonant transition of an oxygen 1s electron into the created hole leading to the peak at 527.0 eV. Notably, the energy difference of the UV-induced peak with respect to the $\pi^*$ resonance of $\approx$4 eV aligns closely with the UV pump photon energy of 4.65 eV. The UV excitation also reduces the intensity of the ground state $\pi^*$ resonance, which indicates that the LUMO is half-filled in the excited state.

\begin{figure*}[ht!]
    \centering
    \includegraphics[width=\textwidth]{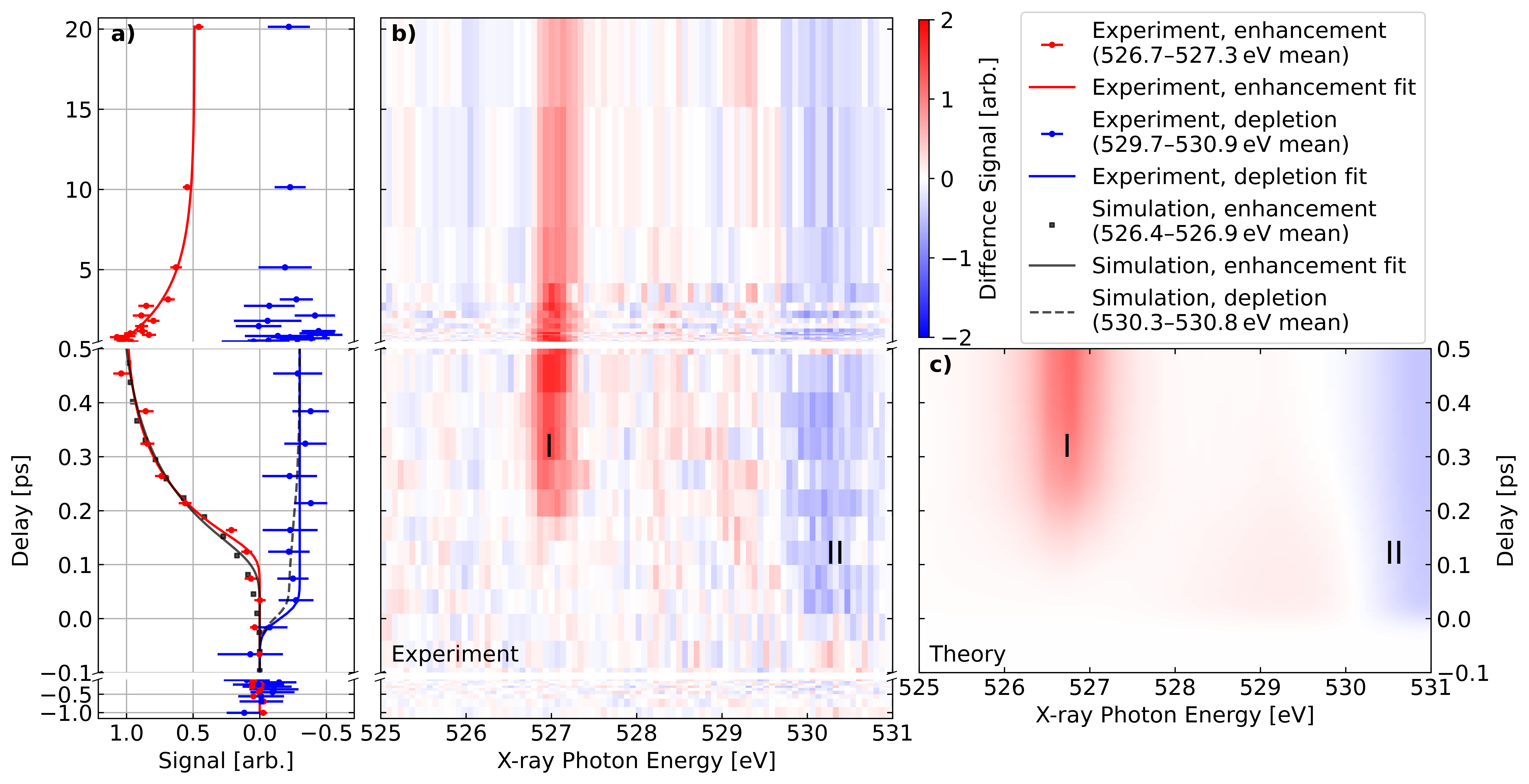}
    \caption{Transient oxygen-edge spectra of AP. a) Comparison of the time evolution of specific regions in the experimental and simulated time-dependent near-edge X-ray absorption fine structure (NEXAFS) difference spectra showing a clear signature of a transient enhancement (I) and depletion (II). b) False-color plot of the experimental time-dependent NEXAFS difference spectra including labels for signatures I and II. c) False-color plot of the simulated time-dependent NEXAFS difference spectra within the first 0.5~ps after UV excitation, based on AIMS simulations. The traces in a) are obtained by averaging the experimental NEXAFS difference signal from panel b) in two energy ranges: (526.7–527.3) eV (red, enhancement) and (529.7–530.9) eV (blue, depletion). The red trace is fitted with a kinetic model over the full 20 ps delay range, while the blue trace is fitted with an error function to capture its onset. The black squares and black dashed line in a) represent averages from the simulated NEXAFS difference spectra from panel c) over (526.4–526.9) eV (enhancement) and (530.3–530.8) eV (depletion), respectively. The time-dependent intensity of the (526.4–526.9) eV in the simulated spectra is fitted by an exponential rise (solid black line). Experimental error bars represent the standard error of the mean, estimated by bootstrapping. The experimental and simulated data are normalized to the corresponding fits (black and red) in panel a).  
    }
    \label{fig:2Dmap}
\end{figure*}

\subsection{Signatures of sub-picosecond dynamics} 
Figure \ref{fig:2Dmap} b) shows the experimental time-dependent difference spectrum of AP for pump-probe delays from -1 to 20~ps resulting from subtracting the static from the UV-excited NEXAFS spectra. For comparison, a simulated spectrum covering time delays up to 0.5~ps after optical excitation is shown in Figure \ref{fig:2Dmap} c). The simulated spectrum is based on the AIMS method,\cite{ben-nun_ab_2000} with the electronic structure calculations based on \textit{hh}-TDA density functional theory (see section \ref{methods} and SI section 5 for details). 

Overall, the simulations show excellent agreement with the experimental results. Both experimental and simulated spectra show the onset of the ground state bleach signature in the 530 - 531~eV region at time zero (II in Fig.~\ref{fig:2Dmap}~b)). In agreement with the experimental data, the most prominent feature in the simulated spectra is a peak red-shifted from the ground state $\pi^*$ resonance. The simulation accurately predicts the energetic position of the transient feature relative to the $\pi^*$ resonance, with a deviation of only 0.3~eV and a slight overestimation of the peak width. The relative intensity of the dominant feature (I in Fig.~\ref{fig:2Dmap}~b)) is also well reproduced (see Fig.~\ref{fig:2Dmap}~a). Moreover, experiment and simulation quantitatively agree on the delay of its temporal onset from the onset of the bleach feature. The simulated difference spectra also show a weak and broad enhancement around 529~eV which appears together with the bleach feature at time zero. Some evidence of a similar feature can also be observed within the first 0.3~ps after time zero. However, it is too close to the noise level to make definitive statements. The reduced signal-to-noise ratio in the depletion region of the experimental difference spectrum arises from the strong ground-state absorption around 531~eV, which both suppresses relative changes due to the large static signal and introduces additional artifacts in the ghost imaging analysis (see the SI section 1 for details).

The onset of the bleach feature marks the experimental time zero, since it corresponds to the loss of some of the ground state population. The weak character of the enhancement which appears at 529~eV co-timed with the bleach feature agrees with expectations based on the $\pi\pi^*$ character of the initially excited state, which exhibits an electron hole in the strongly delocalized $\pi$ orbital (see the orbital visualization in figure \ref{fig:ExpSetup}). Such a hole generally yields weak NEXAFS cross-sections due to the limited overlap with the strongly localized O 1s orbital. The simulation confirms this assignment (see SI section 6). Notably, the peak at 529 eV does not correspond to a $1s \to \pi$ transition but rather originates from excitation into a higher-lying unoccupied orbital. The $1s \to \pi$ transition is located at 525.5~eV in the simulations and even weaker. Thus, it is not discernible in the experimental spectra (see a plot of the simulated $^1\pi\pi^*$ signature in the SI, section 7).

Accordingly, the much stronger peak at 527 eV (526.7 eV in the simulation) must correspond to an excited state with different diabatic character. The higher intensity of this resonance can be explained by an excitation of the oxygen 1s electron into a valence hole with much stronger localization at the oxygen atom. Such a strong localization is typical for oxygen lone pair orbitals (see visualization in Figure \ref{fig:ExpSetup}), i.e., for an $n\pi^*$ excited state character. This assignment is confirmed by the simulations (see SI section 6). Moreover, they provide crucial information about its multiplicity and the nature of the population transfer process. The 526.7~eV peak is associated with the $^1n\pi^*$ state populated by nonadiabatic dynamics through a $^1\pi\pi^*/^1n\pi^*$ CI. 

\begin{figure}[!ht]
    \centering
    \includegraphics[width=0.5\textwidth]{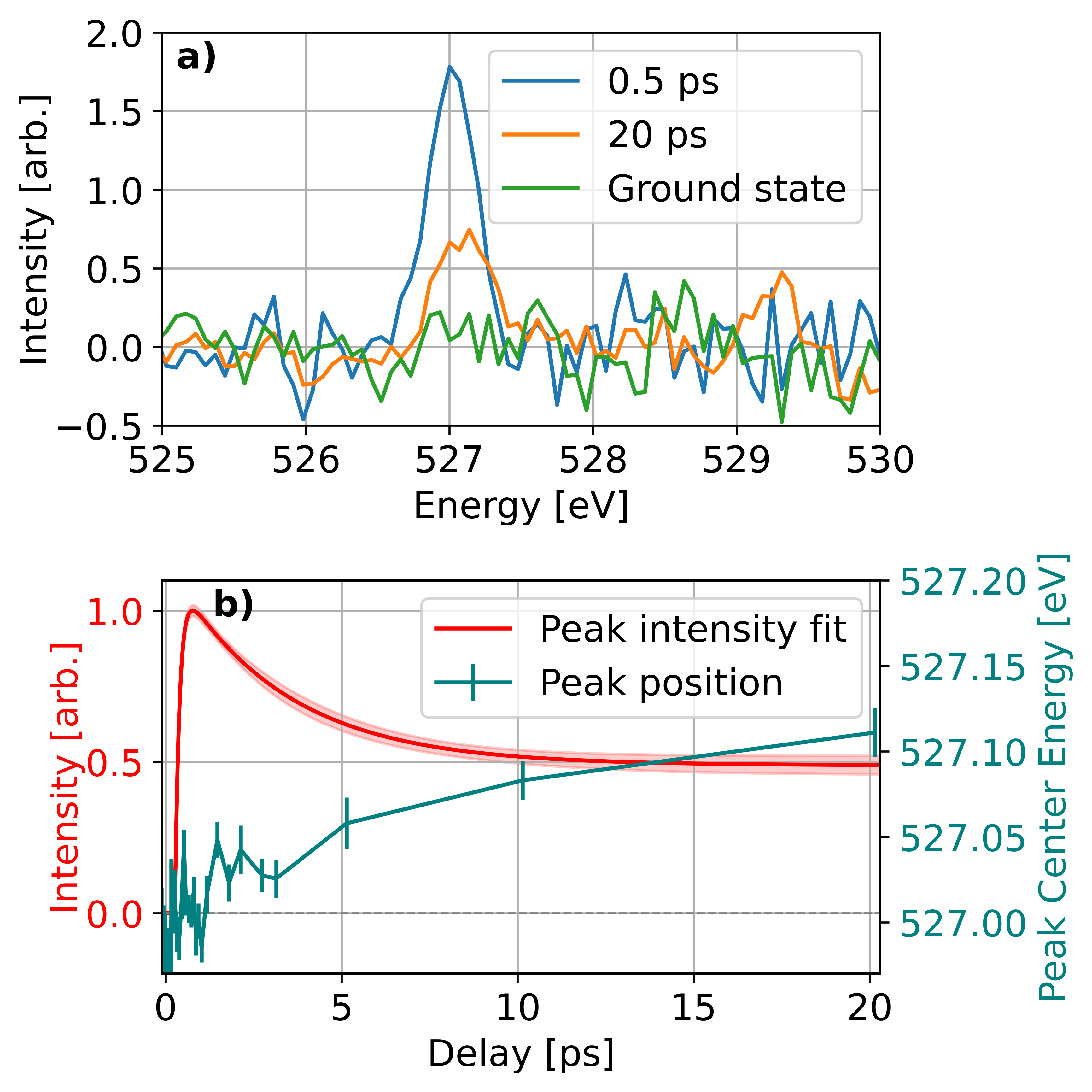}
    \caption{Shift in the spectral position of the $n\pi^{*}$ feature. a) Transient NEXAFS spectra before time zero (green), at 0.5~ps where the $n\pi^*$ feature reaches its maximum intensity, and at 20~ps (orange). b) Comparison between the intensity fit (red, see Fig.~\ref{fig:2Dmap} and equation \ref{eq:fit_base}) and spectral position of the $n\pi^{*}$ transient signature (teal). The spectral position is evaluated by fitting a Gaussian function to the peak. The red shaded area and the teal error bars represent the standard deviation for the respective observable obtained by bootstrapping. }
    \label{fig:PeakShift}
\end{figure}

\subsection{Signatures of picosecond dynamics} 
Both the $n\pi^*$ peak and the ground state bleach are observable throughout the entire experimentally investigated time window, suggesting that within this window the population does not completely return to the ground state. The $n\pi^*$ peak intensity decays on the picosecond timescale and settles into a persistent non-zero plateau within the experimental delay range. Figure \ref{fig:2Dmap} b) indicates a shift of the $n\pi^*$ signature to higher photon energies at long delay steps. This shift is more clearly visible considering spectra at different delay steps as shown in Figure \ref{fig:PeakShift} a). The blue line corresponds to a delay of 0.5~ps, where the $n\pi^{*}$ signature has the highest intensity, the orange line represents the latest delay step at 20~ps. As a baseline reference the green line corresponds to the difference spectrum before time zero. Additionally, the time-dependent peak intensity and position is plotted in Fig.~\ref{fig:PeakShift} b). The peak center shifts by approximately 0.1 eV to higher photon energies while the peak width stays approximately constant. The majority of the peak shift happens after 2.5 ps, after the initial peak intensity decay is complete. 

\subsection{Quantification by kinetic model} 
The delay in onset between the ground state bleach and the $n\pi^*$ peak, which is observed in both experimental and simulated spectra, corresponds to a delay in the onset of $^1\pi\pi^*$ state depopulation (see SI section 4). The delay is caused by the nuclear wavepacket, which is launched by photoexcitation in the Franck-Condon region of the $^1\pi\pi^*$ state, requiring an induction period $t_i$ to reach the $^1\pi\pi^*/^1n\pi^*$ CI seam and to enable efficient population transfer to the $^1n\pi^*$ state. Thus, the population dynamics between $^1\pi\pi^*$ state and $^1n\pi^*$ state cannot be accurately described by a purely exponential model. Accordingly, we quantify the population dynamics underlying the time-dependent intensity changes of the $n\pi^*$ peak $I\left(t\right)$ with the kinetic model of a delayed two stage consecutive process given by (see the SI section 4 for details),

\begin{equation}
    I(t) = \mathrm{IRF}(t) \otimes \Big[ H(t-t_i)\, f(t) \Big]
    \label{eq:fit_base}
\end{equation}

\begin{equation}
    \begin{split}
        f(t) &= I_0 \,\frac{\tau_2}{\tau_2 - \tau_1} 
        \left( e^{-\tfrac{t-t_i}{\tau_2}} - e^{-\tfrac{t-t_i}{\tau_1}} \right) \\
        &\quad + \eta i_0 \left( 1 - \frac{\tau_2}{\tau_2 - \tau_1} e^{-\tfrac{t-t_i}{\tau_2}}
        + \frac{\tau_1}{\tau_2 - \tau_1} e^{-\tfrac{t-t_i}{\tau_1}} \right)
    \end{split}
    \label{eq:fit_func}
\end{equation}

where $A$ is the signal amplitude, $t_i$ the induction period, $\mathrm{IRF}\left(t\right)$ the Gaussian instrument response function with a FWHM of 50~fs, and $\mathrm{H}\left(t-t_i\right)$ the Heaviside step function. The first part of the equation $f\left(t\right)$ describes the population of the $^1n\pi^*$ state with a cross section modulated intensity $I_0$ and the second part describes its depopulation towards an additional state with an absorption cross section that differs by a factor $\eta$. The exponential rise and subsequent decay of the $n\pi^*$ peak are quantified by the time constants $\tau_1 = \left(0.13 \pm 0.02\right)$ ps and $\tau_2 = \left(3.17 \pm 0.66\right)$ ps, respectively. The induction period relative to time zero is quantified as $t_i = (0.12 \pm 0.02)$ ps. Time zero is separately obtained from an error function fit of the ground state bleach onset. A fit with an exponential rise $f(t)=I_0\left(1 - e^{-(t - t_i)/\tau_1}\right)$ of the simulated spectra yields an induction period $t_i = \left(0.102 \pm 0.001\right)$ ps and a time constant $\tau_1 = \left(0.149 \pm 0.002\right)$ ps in quantitative agreement with the experimental fit. The fits are visualized in figure \ref{fig:2Dmap} a). The extracted value of $\eta = \left(0.45 \pm 0.03\right)$ indicates that the effective cross section of the additional state, which the $^1n\pi^*$ state is depopulated to, amounts to only 45\% of that of the $^1n\pi^*$ state.

\subsection{Discussion}

The TR-NEXAFS spectra presented here confirm the fast population of the $^1n\pi^{*}$ state through a directly accessible CI. The spectra reveal an induction period of ($0.12 \pm 0.02$) ps for the $^1n\pi^{*}$ signature. During this period the population resides in the initially populated $^1\pi\pi^{*}$ state that only shows a weak signature. The low signal level is expected based on the electronic character of the state and the insensitivity of the experimental observable to delocalized valence holes. The population transfers through the CI to the $^1n\pi^{*}$ state with a time constant of ($0.13 \pm 0.02$) ps, as evident from the strong signal at 527.0 eV in the NEXAFS spectra. The combination of induction period and decay time constant is by a factor of $\approx$2 larger than the time constant which was identified by Lee \textit{et al.} for the $^1\pi\pi^*$ state depopulation in AP from time-resolved photoelectron spectra.\cite{lee_substituent_2002} The difference can be explained by the use of a longer excitation wavelength (283 nm) in the previous study and a potential windowing effect in the photoelectron spectra due to a strong increase of the ionization potential on the path to the $^1\pi\pi^*/^1n\pi^*$ CI.

After reaching a maximum, the signal decreases with a time constant of ($3.17 \pm 0.66$)~ps. Within the investigated window of 20~ps the signal decreases to a level consistent with the reduced cross section determined by $\eta = \left(0.45 \pm 0.03\right)$, accompanied by a slight energy shift. The magnitude of the drop in intensity can in principle arise from three possible processes, or a combination thereof: structural relaxation within the $^1n\pi^*$ state, its depopulation towards the ground state, or its depopulation towards the triplet manifold via ISC. Structural relaxation is highly unlikely to yield a reduction in cross-section by the observed magnitude by itself, since the cross-sections of NEXAFS transitions are typically not very sensitive to nuclear geometry.\cite{wolf_probing_2017, wolf_transient_2021} Similarly, ground state recovery cannot be the dominating process leading to the observed intensity reduction, since a return to the ground state would be observable by a reduction in the intensity of the bleach region between 530 and 531~eV even at the limited signal-to-noise ratio of the experiment. However, a triplet state with $n\pi^*$ character could very well have a cross-section which is lowered by the observed magnitude. The assignment of the plateaued signal to a $^3n\pi^*$ state would also be consistent with previous observations of ISC with unity quantum yield.\cite{berger_photochemical_1975, zhao_laser_1997} Moreover, previous multi configuration time dependent Hartree simulations starting from the $^1n\pi^*$ state under the approximation of a constant spin-orbit coupling value have predicted fast and efficient intersystem crossing from the $^1n\pi^{*}$ state of AP, albeit at even shorter timescales.\cite{huix-rotllant_population_2016}

To further test the assignment of the intensity decay to ISC to a $^3n\pi^*$ state, we perform OO-DFT calculations, which predict the $^3n\pi^*$ state to have a relative cross-section of 0.56 with respect to the $^1n\pi^*$ state in good agreement of the value of $\eta = 0.45$ obtained from the fit. Additionally, they predict the energy of the O 1s$\rightarrow$n resonance to be essentially the same within the accuracy of the method (shifted by +0.01 eV with respect to the $^1n\pi^*$ state). Thus, the calculations do not contradict the experimentally observed slight shift in photon energy of the $n\pi^*$ resonance coinciding with the stabilization of the signal intensity for higher delays. The difference in relative intensity values from the fit to the experimental data and the calculations can be explained by the accuracy of absorption cross-section predictions by the OO-DFT method. The close agreement indicates that the predominant fraction of the excited population relaxes via this channel. Additionally, a minor ground state recovery channel and effects from a weak dependence of the resonance intensity on the nuclear structure could contribute. 

The $^3n\pi^*$ state is unlikely to be populated by a direct transition from the $^1n\pi^{*}$ state. Such a transition is forbidden by the El Sayed selection rule.\cite{el_sayed_spin-orbit_1963} Therefore, it would not fit the observed timescale of $^3n\pi^*$ state population. Instead, it is likely mediated by a very short lived $^3\pi\pi^{*}$ state, which the population passes before reaching the $^3n\pi^{*}$ state (see Figure \ref{fig:ExpSetup}c)). Due to the spin-forbidden nature of ISC between the $^1n\pi^*$ and the $^3\pi\pi^*$ state, it can be expected to be significantly less efficient than the consecutive spin-allowed IC between $^3\pi\pi^*$ and $^3n\pi^*$ states. Thus, ISC can be expected to be the rate-limiting step and the population residing in the $^3\pi\pi^*$ state to be infinitesimally low at any given point in time. Additionally, its NEXAFS cross-section can be expected to be low due to its delocalized character, analogous to the signature of the $^1\pi\pi^*$ state. As a result the TR-NEXAFS spectra do not show a signature of this intermediate state. 

Our findings are in contrast to the results from previous ultrafast electron diffraction studies.\cite{feenstra_excited_2005, park_ultrafast_2006}. These identified structural signatures pointing to a bifurcation of pathways with part of the population undergoing a Norrish type I reaction in the $^1n\pi^*$ state and the remaining population undergoing ISC to a $^3\pi\pi^*$ state. Based on our results, we cannot rule out a minor fraction of the population undergoing a Norrish Type I reaction from the singlet manifold. Such a minor channel could be more prominent in the diffraction observable due to the associated large structural change. The evidence for the electronic character of the triplet state from UED is only indirect based on the refined structure. In contrast, our results provide direct evidence for the population of a $^3n\pi^*$ state as the dominant channel.

\section{Conclusion}
Our investigation of the excited state dynamics of AP using a combination of time-resolved NEXAFS spectroscopy at the oxygen edge and AIMS simulations at the \textit{hh}-TDA DFT level showcase the ability of the NEXAFS observable to follow excited state population dynamics across states of various electronic character and multiplicity. For the case of AP, it reveals the mechanism of triplet state population through IC from the photoexcited $^1\pi\pi^*$ state through a $^1\pi\pi^*/^1n\pi^*$ conical intersection followed by ISC to a $^3\pi\pi^*$ state, which is in turn followed by IC to a $^3n\pi^*$ state. Moreover, our study unambiguously identifies the $^3n\pi^*$ state as the photochemically active state for the Norrish type I reaction in AP. The identified population mechanism of the triplet manifold can likely be generalized to any aromatic carbonyl species. We find close to quantitative agreement between our experimental results and the AIMS simulations of the dynamics including the NEXAFS observable in the singlet manifold with the \textit{hh}-TDA method showcasing its predictive power for investigations of excited state dynamics involving states with diverse diabatic character.

\section*{Acknowledgements}
This work was supported by the AMOS program within the U.S. Department of Energy, Office of Science, Basic Energy Sciences, Chemical Sciences, Geosciences, and Biosciences Division under Contract No. DE-AC02-76SF00515. P.U. acknowledges support by the National Science Foundation MPS-Ascend Postdoctoral Research Fellowship, under Grant No. 2213324. A.E.H.B. acknowledges financial support from the Novo Nordisk Foundation under grant reference number NNF24OC0089345. D.H. was a Stanford Science Fellow for the initial stages of this work. The Flatiron Institute is a division of the Simons Foundation. A.E.G. was supported by the European Union, through Horizon Europe Project No. 123-CO: 101067645. Views and opinions expressed are, however, those of the authors only and do not necessarily reflect those of the European Union. Neither the European Union nor the granting authority can be held responsible for them. R.F. was supported by the U.S. Department of Energy, Office of Science, Office of Basic Energy Sciences Award No. DE-SC0026078. Use of the Linac Coherent Light Source (LCLS), SLAC National Accelerator Laboratory, is supported by the U.S. Department of Energy, Office of Science, Office of Basic Energy Sciences under Contract No. DE-AC02-76SF00515.

\printbibliography

@book{dinda_essentials_2017,
	address = {Cham},
	series = {Lecture {Notes} in {Chemistry}},
	title = {Essentials of {Pericyclic} and {Photochemical} {Reactions}},
	isbn = {978-3-319-45934-9},
	doi = {10.1007/978-3-319-45934-9},
	language = {eng},
	number = {93},
	publisher = {Springer},
	author = {Dinda, Biswanath},
	year = {2017},
}

@book{padwa_organic_2017,
	address = {Boca Raton},
	title = {Organic {Photochemistry}},
	isbn = {978-0-203-74482-6},
	doi = {10.1201/9780203744826},
	publisher = {CRC Press},
	editor = {Padwa, Albert},
	month = oct,
	year = {2017},
}

@article{li_ghost-imaging-enhanced_2022,
	title = {Ghost-imaging-enhanced noninvasive spectral characterization of stochastic x-ray free-electron-laser pulses},
	volume = {5},
	copyright = {2022 UChicago Argonne, LLC, Operator of Argonne National Laboratory},
	issn = {2399-3650},
	url = {https://www.nature.com/articles/s42005-022-00962-8},
	doi = {10.1038/s42005-022-00962-8},
	language = {en},
	number = {1},
	urldate = {2026-01-05},
	journal = {Communications Physics},
	author = {Li, Kai and Laksman, Joakim and Mazza, Tommaso and Doumy, Gilles and Koulentianos, Dimitris and Picchiotti, Alessandra and Serkez, Svitozar and Rohringer, Nina and Ilchen, Markus and Meyer, Michael and Young, Linda},
	month = jul,
	year = {2022},
	note = {Publisher: Nature Publishing Group},
	pages = {191},
}

@article{wight_k-shell_1974,
	title = {K-{Shell} energy loss spectra of 2.5 {keV} electrons in {CO2} and {N2O}},
	volume = {3},
	issn = {0368-2048},
	url = {http://www.sciencedirect.com/science/article/pii/0368204874800101},
	doi = {10.1016/0368-2048(74)80010-1},
	language = {en},
	number = {3},
	urldate = {2019-12-15},
	journal = {Journal of Electron Spectroscopy and Related Phenomena},
	author = {Wight, G. R. and Brion, C. E.},
	month = jan,
	year = {1974},
	pages = {191--205},
}

@article{walter_time-resolved_2022,
	title = {The time-resolved atomic, molecular and optical science instrument at the {Linac} {Coherent} {Light} {Source}},
	volume = {29},
	copyright = {https://creativecommons.org/licenses/by/4.0/},
	issn = {1600-5775},
	url = {https://journals.iucr.org/s/issues/2022/04/00/gb5129/},
	doi = {10.1107/S1600577522004283},
	number = {4},
	urldate = {2022-07-10},
	journal = {Journal of Synchrotron Radiation},
	author = {Walter, P. and Osipov, T. and Lin, M.-F. and Cryan, J. and Driver, T. and Kamalov, A. and Marinelli, A. and Robinson, J. and Seaberg, M. H. and Wolf, T. J. A. and Aldrich, J. and Brown, N. and Champenois, E. G. and Cheng, X. and Cocco, D. and Conder, A. and Curiel, I. and Egger, A. and Glownia, J. M. and Heimann, P. and Holmes, M. and Johnson, T. and Lee, L. and Li, X. and Moeller, S. and Morton, D. S. and Ng, M. L. and Ninh, K. and O'Neal, J. T. and Obaid, R. and Pai, A. and Schlotter, W. and Shepard, J. and Shivaram, N. and Stefan, P. and Van, X. and Wang, A. L. and Wang, H. and Yin, J. and Yunus, S. and Fritz, D. and James, J. and Castagna, J.-C.},
	month = jul,
	year = {2022},
	note = {Number: 4
Publisher: International Union of Crystallography},
	pages = {957--968},
}

@article{hitchcock_inner-shell_1992,
    title = {Inner-shell spectroscopy of benzaldehyde, terephthalaldehyde, ethylbenzoate, terephthaloyl chloride and phosgene: models for core excitation of poly(ethylene terephthalate)},
    volume = {96},
    issn = {0022-3654, 1541-5740},
    shorttitle = {Inner-shell spectroscopy of benzaldehyde, terephthalaldehyde, ethylbenzoate, terephthaloyl chloride and phosgene},
    url = {https://pubs.acs.org/doi/abs/10.1021/j100201a015},
    doi = {10.1021/j100201a015},
    language = {en},
    number = {22},
    urldate = {2025-05-09},
    journal = {The Journal of Physical Chemistry},
    author = {Hitchcock, A. P. and Urquhart, S. G. and Rightor, E. G.},
    month = oct,
    year = {1992},
    pages = {8736--8750},
}

@article{borne_design_2024,
    title = {Design and performance of a magnetic bottle electron spectrometer for high-energy photoelectron spectroscopy},
    volume = {95},
    issn = {0034-6748},
    url = {https://doi.org/10.1063/5.0223334},
    doi = {10.1063/5.0223334},
    number = {12},
    urldate = {2025-05-09},
    journal = {Review of Scientific Instruments},
    author = {Borne, Kurtis and O’Neal, Jordan T. and Wang, Jun and Isele, Erik and Obaid, Razib and Berrah, Nora and Cheng, Xinxin and Bucksbaum, Philip H. and James, Justin and Kamalov, Andrei and Larsen, Kirk A. and Li, Xiang and Lin, Ming-Fu and Liu, Yusong and Marinelli, Agostino and Summers, Adam M. and Thierstein, Emily and Wolf, Thomas J. A. and Rolles, Daniel and Walter, Peter and Cryan, James P. and Driver, Taran},
    month = dec,
    year = {2024},
    pages = {125110},
}

@article{droste_high-sensitivity_2020,
    title = {High-sensitivity x-ray/optical cross-correlator for next generation free-electron lasers},
    volume = {28},
    copyright = {© 2020 Optical Society of America},
    issn = {1094-4087},
    url = {https://opg.optica.org/oe/abstract.cfm?uri=oe-28-16-23545},
    doi = {10.1364/OE.398048},
    language = {EN},
    number = {16},
    urldate = {2025-05-12},
    journal = {Optics Express},
    author = {Droste, Stefan and Zohar, Sioan and Shen, Lingjia and White, Vaughn E. and Diaz-Jacobo, Elizabeth and Coffee, Ryan N. and Reid, Alexander H. and Tavella, Franz and Minitti, Michael P. and Turner, Joshua J. and Robinson, Joseph S. and Fry, Alan R. and Coslovich, Giacomo},
    month = aug,
    year = {2020},
    note = {Publisher: Optica Publishing Group},
    pages = {23545--23553},
}

@article{li_time-resolved_2021,
    title = {Time-resolved pump–probe spectroscopy with spectral domain ghost imaging},
    volume = {228},
    issn = {1364-5498},
    url = {https://pubs.rsc.org/en/content/articlelanding/2021/fd/d0fd00122h},
    doi = {10.1039/D0FD00122H},
    language = {en},
    number = {0},
    urldate = {2025-05-12},
    journal = {Faraday Discussions},
    author = {Li, Siqi and Driver, Taran and Alexander, Oliver and Cooper, Bridgette and Garratt, Douglas and Marinelli, Agostino and Cryan, James P. and Marangos, Jonathan P.},
    month = may,
    year = {2021},
    note = {Publisher: The Royal Society of Chemistry},
    pages = {488--501},
}

@article{li_two-dimensional_2021,
    title = {Two-dimensional correlation analysis for x-ray photoelectron spectroscopy},
    volume = {54},
    issn = {0953-4075},
    url = {https://dx.doi.org/10.1088/1361-6455/abcdf1},
    doi = {10.1088/1361-6455/abcdf1},
    language = {en},
    number = {14},
    urldate = {2025-05-12},
    journal = {Journal of Physics B: Atomic, Molecular and Optical Physics},
    author = {Li, S and Driver, T and Al Haddad, A and Champenois, E G and Agåker, M and Alexander, O and Barillot, T and Bostedt, C and Garratt, D and Kjellsson, L and Lutman, A A and Rubensson, J-E and Sathe, C and Marinelli, A and Marangos, J P and Cryan, J P},
    month = aug,
    year = {2021},
    note = {Publisher: IOP Publishing},
    pages = {144005},
}

@article{warren_s2_1986,
	title = {The {S2}$\leftarrow${S0} laser photoexcitation spectrum and excited state dynamics of jet‐cooled acetophenone},
	volume = {85},
	issn = {0021-9606},
	url = {https://aip-scitation-org.stanford.idm.oclc.org/doi/abs/10.1063/1.451090},
	doi = {10.1063/1.451090},
	number = {5},
	urldate = {2022-05-04},
	journal = {The Journal of Chemical Physics},
	author = {Warren, J. A. and Bernstein, E. R.},
	month = sep,
	year = {1986},
	note = {Publisher: American Institute of Physics},
	pages = {2365--2367},
}

@article{bao_recent_2022,
    title = {Recent {Trends} in {Advanced} {Photoinitiators} for {Vat} {Photopolymerization} {3D} {Printing}},
    volume = {43},
    copyright = {© 2022 The Authors. Macromolecular Rapid Communications published by Wiley-VCH GmbH},
    issn = {1521-3927},
    url = {https://onlinelibrary.wiley.com/doi/abs/10.1002/marc.202200202},
    doi = {10.1002/marc.202200202},
    language = {en},
    number = {14},
    urldate = {2025-07-18},
    journal = {Macromolecular Rapid Communications},
    author = {Bao, Yinyin},
    year = {2022},
    note = {\_eprint: https://onlinelibrary.wiley.com/doi/pdf/10.1002/marc.202200202},
    pages = {2200202},
}

@article{wolf_probing_2017,
	title = {Probing ultrafast ππ*/nπ* internal conversion in organic chromophores via {K}-edge resonant absorption},
	volume = {8},
	copyright = {2017 The Author(s)},
	issn = {2041-1723},
	url = {https://www.nature.com/articles/s41467-017-00069-7},
	doi = {10.1038/s41467-017-00069-7},
	language = {en},
	number = {1},
	urldate = {2024-02-15},
	journal = {Nature Communications},
	author = {Wolf, T. J. A. and Myhre, R. H. and Cryan, J. P. and Coriani, S. and Squibb, R. J. and Battistoni, A. and Berrah, N. and Bostedt, C. and Bucksbaum, P. and Coslovich, G. and Feifel, R. and Gaffney, K. J. and Grilj, J. and Martinez, T. J. and Miyabe, S. and Moeller, S. P. and Mucke, M. and Natan, A. and Obaid, R. and Osipov, T. and Plekan, O. and Wang, S. and Koch, H. and Gühr, M.},
	month = jun,
	year = {2017},
	note = {Number: 1
Publisher: Nature Publishing Group},
	pages = {29},
}

@article{kjonstad_photoinduced_2024,
	title = {Photoinduced hydrogen dissociation in thymine predicted by coupled cluster theory},
	volume = {15},
	issn = {2041-1723},
	url = {https://www.ncbi.nlm.nih.gov/pmc/articles/PMC11584849/},
	doi = {10.1038/s41467-024-54436-2},
	urldate = {2025-01-30},
	journal = {Nature Communications},
	author = {Kjønstad, Eirik F. and Fajen, O. Jonathan and Paul, Alexander C. and Angelico, Sara and Mayer, Dennis and Gühr, Markus and Wolf, Thomas J. A. and Martínez, Todd J. and Koch, Henrik},
	month = nov,
	year = {2024},
	pmid = {39578441},
	pmcid = {PMC11584849},
	pages = {10128},
}

@article{karkas_photochemical_2016,
	title = {Photochemical {Approaches} to {Complex} {Chemotypes}: {Applications} in {Natural} {Product} {Synthesis}},
	volume = {116},
	issn = {0009-2665},
	shorttitle = {Photochemical {Approaches} to {Complex} {Chemotypes}},
	url = {https://doi.org/10.1021/acs.chemrev.5b00760},
	doi = {10.1021/acs.chemrev.5b00760},
	number = {17},
	urldate = {2025-07-18},
	journal = {Chemical Reviews},
	author = {Kärkäs, Markus D. and Porco, John A. Jr. and Stephenson, Corey R. J.},
	month = sep,
	year = {2016},
	note = {Publisher: American Chemical Society},
	pages = {9683--9747},
}

@article{muller_recent_2022,
	title = {Recent {Advances} in {Type} {I} {Photoinitiators} for {Visible} {Light} {Induced} {Photopolymerization}},
	volume = {6},
	copyright = {© 2022 The Authors. ChemPhotoChem published by Wiley-VCH GmbH},
	issn = {2367-0932},
	url = {https://onlinelibrary.wiley.com/doi/abs/10.1002/cptc.202200091},
	doi = {10.1002/cptc.202200091},
	language = {en},
	number = {11},
	urldate = {2025-07-20},
	journal = {ChemPhotoChem},
	author = {Müller, Stefanie Monika and Schlögl, Sandra and Wiesner, Tanja and Haas, Michael and Griesser, Thomas},
	year = {2022},
	note = {\_eprint: https://chemistry-europe.onlinelibrary.wiley.com/doi/pdf/10.1002/cptc.202200091},
	pages = {e202200091},
}

@article{bannwarth_holehole_2020,
	title = {Hole–hole {Tamm}–{Dancoff}-approximated density functional theory: {A} highly efficient electronic structure method incorporating dynamic and static correlation},
	volume = {153},
	issn = {0021-9606},
	shorttitle = {Hole–hole {Tamm}–{Dancoff}-approximated density functional theory},
	url = {https://doi.org/10.1063/5.0003985},
	doi = {10.1063/5.0003985},
	number = {2},
	urldate = {2024-09-14},
	journal = {The Journal of Chemical Physics},
	author = {Bannwarth, Christoph and Yu, Jimmy K. and Hohenstein, Edward G. and Martínez, Todd J.},
	month = jul,
	year = {2020},
	pages = {024110},
}

@article{yu2020ab,
  title={Ab initio nonadiabatic molecular dynamics with hole--hole Tamm--Dancoff approximated density functional theory},
  author={Yu, Jimmy K and Bannwarth, Christoph and Hohenstein, Edward G and Mart{\'\i}nez, Todd J},
  journal={Journal of Chemical Theory and Computation},
  volume={16},
  number={9},
  pages={5499--5511},
  year={2020},
  publisher={ACS Publications}
}

@article{weigend2005balanced,
  title={Balanced basis sets of split valence, triple zeta valence and quadruple zeta valence quality for H to Rn: Design and assessment of accuracy},
  author={Weigend, Florian and Ahlrichs, Reinhart},
  journal={Physical Chemistry Chemical Physics},
  volume={7},
  number={18},
  pages={3297--3305},
  year={2005},
  publisher={Royal Society of Chemistry}
}

@article{seritan2021terachem,
  title={TeraChem: A graphical processing unit-accelerated electronic structure package for large-scale ab initio molecular dynamics},
  author={Seritan, Stefan and Bannwarth, Christoph and Fales, Bryan S and Hohenstein, Edward G and Isborn, Christine M and Kokkila-Schumacher, Sara IL and Li, Xin and Liu, Fang and Luehr, Nathan and Snyder Jr, James W and others},
  journal={Wiley Interdisciplinary Reviews: Computational Molecular Science},
  volume={11},
  number={2},
  pages={e1494},
  year={2021},
  publisher={Wiley Online Library}
}

@article{seritan2020terachem,
  title={TeraChem: Accelerating electronic structure and ab initio molecular dynamics with graphical processing units},
  author={Seritan, Stefan and Bannwarth, Christoph and Fales, B Scott and Hohenstein, Edward G and Kokkila-Schumacher, Sara IL and Luehr, Nathan and Snyder, James W and Song, Chenchen and Titov, Alexey V and Ufimtsev, Ivan S and others},
  journal={The Journal of chemical physics},
  volume={152},
  number={22},
  year={2020},
  pages={224110},
  publisher={AIP Publishing}
}

@article{becke1988density,
  title={Density-functional exchange-energy approximation with correct asymptotic behavior},
  author={Becke, Axel D},
  journal={Physical review A},
  volume={38},
  number={6},
  pages={3098},
  year={1988},
  publisher={APS}
}

@article{lee1988development,
  title={Development of the Colle-Salvetti correlation-energy formula into a functional of the electron density},
  author={Lee, Chengteh and Yang, Weitao and Parr, Robert G},
  journal={Physical review B},
  volume={37},
  number={2},
  pages={785},
  year={1988},
  publisher={APS}
}

@article{becke1993new,
  title={A new mixing of Hartree-Fock and local density-functional theories},
  author={Becke, Axel D},
  journal={Journal of chemical Physics},
  volume={98},
  number={2},
  pages={1372--1377},
  year={1993}
}

@article{el_sayed_spin-orbit_1963,
	title = {Spin-orbit coupling and the radiationless processes in nitrogen heterocycles},
	volume = {38},
	doi = {10.1063/1.1733610},
	journal = {J. Chem. Phys.},
	author = {El Sayed, M. A.},
	year = {1963},
	pages = {2834--2838},
}

@article{rueggeberg_state---art_2011,
	title = {State-of-the-art: {Dental} photocuring {A} review},
	volume = {27},
	issn = {0109-5641},
	url = {http://www.sciencedirect.com/science/article/pii/S0109564110004641},
	doi = {10.1016/j.dental.2010.10.021},
	number = {1},
	journal = {Dent. Mater.},
	author = {Rueggeberg, Frederick A.},
	year = {2011},
	pages = {39},
}

@article{berger_photochemical_1975,
	title = {Photochemical and photophysical processes in acetophenone},
	volume = {97},
	issn = {0002-7863, 1520-5126},
	url = {https://pubs.acs.org/doi/abs/10.1021/ja00850a006},
	doi = {10.1021/ja00850a006},
	language = {en},
	number = {17},
	urldate = {2024-07-03},
	journal = {Journal of the American Chemical Society},
	author = {Berger, Michael and Steel, Colin},
	month = aug,
	year = {1975},
	pages = {4817--4821},
}

@article{zhao_laser_1997,
	title = {A laser photofragmentation time-of-flight mass spectrometric study of acetophenone at 193 and 248 nm},
	volume = {107},
	issn = {0021-9606},
	url = {https://doi.org/10.1063/1.474964},
	doi = {10.1063/1.474964},
	number = {18},
	urldate = {2024-07-03},
	journal = {The Journal of Chemical Physics},
	author = {Zhao, H.-Q. and Cheung, Y.-S. and Liao, C.-L. and Liao, C.-X. and Ng, C. Y. and Li, Wai-Kee},
	month = nov,
	year = {1997},
	pages = {7230--7241},
}

@article{lee_substituent_2002,
	title = {Substituent {Effects} in {Molecular} {Electronic} {Relaxation} {Dynamics} via {Time}-{Resolved} {Photoelectron} {Spectroscopy}:  ππ* {States} in {Benzenes}},
	volume = {106},
	issn = {1089-5639},
	shorttitle = {Substituent {Effects} in {Molecular} {Electronic} {Relaxation} {Dynamics} via {Time}-{Resolved} {Photoelectron} {Spectroscopy}},
	url = {https://doi.org/10.1021/jp021096h},
	doi = {10.1021/jp021096h},
	number = {39},
	urldate = {2024-07-03},
	journal = {The Journal of Physical Chemistry A},
	author = {Lee, Shih-Huang and Tang, Kuo-Chun and Chen, I-Chia and Schmitt, M. and Shaffer, J. P. and Schultz, T. and Underwood, Jonathan G. and Zgierski, M. Z. and Stolow, Albert},
	month = oct,
	year = {2002},
	note = {Publisher: American Chemical Society},
	pages = {8979--8991},
}

@article{park_ultrafast_2006,
	title = {Ultrafast electron diffraction: {Excited} state structures and chemistries of aromatic carbonyls},
	volume = {124},
	issn = {0021-9606},
	shorttitle = {Ultrafast electron diffraction},
	url = {https://doi.org/10.1063/1.2194017},
	doi = {10.1063/1.2194017},
	number = {17},
	urldate = {2024-07-03},
	journal = {The Journal of Chemical Physics},
	author = {Park, Sang Tae and Feenstra, Jonathan S. and Zewail, Ahmed H.},
	month = may,
	year = {2006},
	pages = {174707},
}

@article{feenstra_excited_2005,
	title = {Excited state molecular structures and reactions directly determined by ultrafast electron diffraction},
	volume = {123},
	issn = {0021-9606},
	url = {https://doi.org/10.1063/1.2140700},
	doi = {10.1063/1.2140700},
	number = {22},
	urldate = {2024-07-03},
	journal = {The Journal of Chemical Physics},
	author = {Feenstra, Jonathan S. and Park, Sang Tae and Zewail, Ahmed H.},
	month = dec,
	year = {2005},
	pages = {221104},
}

@article{larsen_compact_2023,
	title = {Compact single-shot soft {X}-ray photon spectrometer for free-electron laser diagnostics},
	volume = {31},
	issn = {1094-4087},
	url = {https://opg.optica.org/abstract.cfm?URI=oe-31-22-35822},
	doi = {10.1364/OE.502105},
	language = {en},
	number = {22},
	urldate = {2023-11-27},
	journal = {Optics Express},
	author = {Larsen, Kirk A. and Borne, Kurtis and Obaid, Razib and Kamalov, Andrei and Liu, Yusong and Cheng, Xinxin and James, Justin and Driver, Taran and Li, Kenan and Liu, Yanwei and Sakdinawat, Anne and David, Christian and Wolf, Thomas J. A. and Cryan, James P. and Walter, Peter and Lin, Ming-Fu},
	month = oct,
	year = {2023},
	pages = {35822},
}

@article{neese_shark_2023,
	title = {The {SHARK} integral generation and digestion system},
	volume = {44},
	copyright = {© 2022 The Author. Journal of Computational Chemistry published by Wiley Periodicals LLC.},
	issn = {1096-987X},
	url = {https://onlinelibrary.wiley.com/doi/abs/10.1002/jcc.26942},
	doi = {10.1002/jcc.26942},
	language = {en},
	number = {3},
	urldate = {2025-07-08},
	journal = {Journal of Computational Chemistry},
	author = {Neese, Frank},
	year = {2023},
	note = {\_eprint: https://onlinelibrary.wiley.com/doi/pdf/10.1002/jcc.26942},
	pages = {381--396},
}

@article{neese_efficient_2009,
	series = {Moving {Frontiers} in {Quantum} {Chemistry}:},
	title = {Efficient, approximate and parallel {Hartree}–{Fock} and hybrid {DFT} calculations. {A} ‘chain-of-spheres’ algorithm for the {Hartree}–{Fock} exchange},
	volume = {356},
	issn = {0301-0104},
	url = {https://www.sciencedirect.com/science/article/pii/S0301010408005089},
	doi = {10.1016/j.chemphys.2008.10.036},
	number = {1},
	urldate = {2025-07-08},
	journal = {Chemical Physics},
	author = {Neese, Frank and Wennmohs, Frank and Hansen, Andreas and Becker, Ute},
	month = feb,
	year = {2009},
	pages = {98--109},
}

@article{neese_improvement_2003,
	title = {An improvement of the resolution of the identity approximation for the formation of the {Coulomb} matrix},
	volume = {24},
	copyright = {Copyright © 2003 Wiley Periodicals, Inc.},
	issn = {1096-987X},
	url = {https://onlinelibrary.wiley.com/doi/abs/10.1002/jcc.10318},
	doi = {10.1002/jcc.10318},
	language = {en},
	number = {14},
	urldate = {2025-07-08},
	journal = {Journal of Computational Chemistry},
	author = {Neese, Frank},
	year = {2003},
	note = {\_eprint: https://onlinelibrary.wiley.com/doi/pdf/10.1002/jcc.10318},
	pages = {1740--1747},
}

@article{neese_efficient_2002,
	title = {Efficient use of the resolution of the identity approximation in time-dependent density functional calculations with hybrid density functionals},
	volume = {362},
	issn = {0009-2614},
	url = {https://www.sciencedirect.com/science/article/pii/S0009261402010539},
	doi = {10.1016/S0009-2614(02)01053-9},
	number = {1},
	urldate = {2025-07-08},
	journal = {Chemical Physics Letters},
	author = {Neese, Frank and Olbrich, Gottfried},
	month = aug,
	year = {2002},
	pages = {170--178},
}

@article{neese_software_2022,
	title = {Software update: {The} {ORCA} program system—{Version} 5.0},
	volume = {12},
	copyright = {© 2022 The Author. WIREs Computational Molecular Science published by Wiley Periodicals LLC.},
	issn = {1759-0884},
	shorttitle = {Software update},
	url = {https://onlinelibrary.wiley.com/doi/abs/10.1002/wcms.1606},
	doi = {10.1002/wcms.1606},
	language = {en},
	number = {5},
	urldate = {2025-07-08},
	journal = {WIREs Computational Molecular Science},
	author = {Neese, Frank},
	year = {2022},
	note = {\_eprint: https://wires.onlinelibrary.wiley.com/doi/pdf/10.1002/wcms.1606},
	pages = {e1606},
}

@article{helmich-paris_improved_2021,
	title = {An improved chain of spheres for exchange algorithm},
	volume = {155},
	issn = {0021-9606},
	url = {https://doi.org/10.1063/5.0058766},
	doi = {10.1063/5.0058766},
	number = {10},
	urldate = {2025-07-08},
	journal = {The Journal of Chemical Physics},
	author = {Helmich-Paris, Benjamin and de Souza, Bernardo and Neese, Frank and Izsák, Róbert},
	month = sep,
	year = {2021},
	pages = {104109},
}

@article{ben1998nonadiabatic,
  title={Nonadiabatic molecular dynamics: Validation of the multiple spawning method for a multidimensional problem},
  author={Ben-Nun, M and Mart{\i}nez, Todd J},
  journal={The Journal of chemical physics},
  volume={108},
  number={17},
  pages={7244--7257},
  year={1998},
  publisher={American Institute of Physics}
}

@article{ben-nun_ab_2000,
	title = {Ab {Initio} {Multiple} {Spawning}: {Photochemistry} from {First} {Principles} {Quantum} {Molecular} {Dynamics}},
	volume = {104},
	url = {http://pubs.acs.org/doi/abs/10.1021/jp994174i},
	doi = {10.1021/jp994174i},
	number = {22},
	journal = {Journal of Physical Chemistry A},
	author = {Ben-Nun, M. and Quenneville, Jason and Martínez, Todd J.},
	year = {2000},
	pages = {5161--5175},
}

@article{wigner1932quantum,
  title={On the quantum correction for thermodynamic equilibrium},
  author={Wigner, Eugene},
  journal={Physical review},
  volume={40},
  number={5},
  pages={749},
  year={1932},
  publisher={APS}
}

@article{epifanovsky2021software,
  title={Software for the frontiers of quantum chemistry: An overview of developments in the Q-Chem 5 package},
  author={Epifanovsky, Evgeny and Gilbert, Andrew TB and Feng, Xintian and Lee, Joonho and Mao, Yuezhi and Mardirossian, Narbe and Pokhilko, Pavel and White, Alec F and Coons, Marc P and Dempwolff, Adrian L and others},
  journal={The Journal of chemical physics},
  volume={155},
  number={8},
  year={2021},
  pages={084801},
  publisher={AIP Publishing}
}

@article{hait2021orbital,
  title={Orbital optimized density functional theory for electronic excited states},
  author={Hait, Diptarka and Head-Gordon, Martin},
  journal={The journal of physical chemistry letters},
  volume={12},
  number={19},
  pages={4517--4529},
  year={2021},
  publisher={ACS Publications}
}

@article{cunha2022relativistic,
  title={Relativistic orbital-optimized density functional theory for accurate core-level spectroscopy},
  author={Cunha, Leonardo A and Hait, Diptarka and Kang, Richard and Mao, Yuezhi and Head-Gordon, Martin},
  journal={The journal of physical chemistry letters},
  volume={13},
  number={15},
  pages={3438--3449},
  year={2022},
  publisher={ACS Publications}
}

@article{sun2015strongly,
  title={Strongly constrained and appropriately normed semilocal density functional},
  author={Sun, Jianwei and Ruzsinszky, Adrienn and Perdew, John P},
  journal={Physical review letters},
  volume={115},
  number={3},
  pages={036402},
  year={2015},
  publisher={APS}
}

@article{jensen2014unifying,
  title={Unifying general and segmented contracted basis sets. Segmented polarization consistent basis sets},
  author={Jensen, Frank},
  journal={Journal of chemical theory and computation},
  volume={10},
  number={3},
  pages={1074--1085},
  year={2014},
  publisher={ACS Publications}
}

@article{ambroise2018probing,
  title={Probing basis set requirements for calculating core ionization and core excitation spectroscopy by the $\Delta$ self-consistent-field approach},
  author={Ambroise, Maximilien A and Jensen, Frank},
  journal={Journal of chemical theory and computation},
  volume={15},
  number={1},
  pages={325--337},
  year={2018},
  publisher={ACS Publications}
}

@article{becke1993density,
  title={Density-functional thermochemistry. III. The role of exact exchange},
  author={Becke, Axel D},
  journal={The Journal of chemical physics},
  volume={98},
  number={7},
  pages={5648--5652},
  year={1993},
  publisher={American Institute of Physics}
}

@article{grimme2010consistent,
  title={A consistent and accurate ab initio parametrization of density functional dispersion correction (DFT-D) for the 94 elements H-Pu},
  author={Grimme, Stefan and Antony, Jens and Ehrlich, Stephan and Krieg, Helge},
  journal={The Journal of chemical physics},
  volume={132},
  number={15},
  year={2010},
  pages={154104},
  publisher={AIP Publishing}
}

@article{ditchfield1971self,
  title={Self-consistent molecular-orbital methods. IX. An extended Gaussian-type basis for molecular-orbital studies of organic molecules},
  author={Ditchfield, RHWJ and Hehre, Warren J and Pople, John A},
  journal={The Journal of Chemical Physics},
  volume={54},
  number={2},
  pages={724--728},
  year={1971},
  publisher={American Institute of Physics}
}

@article{cederbaum1980many,
  title={Many-body theory of core holes},
  author={Cederbaum, Lorenz S and Domcke, W and Schirmer, Jochen},
  journal={Physical Review A},
  volume={22},
  number={1},
  pages={206},
  year={1980},
  publisher={APS}
}

@article{huix-rotllant_population_2016,
	title = {Population of triplet states in acetophenone: {A} quantum dynamics perspective},
	volume = {19},
	issn = {1878-1543},
	shorttitle = {Population of triplet states in acetophenone},
	url = {https://comptes-rendus.academie-sciences.fr/chimie/articles/en/10.1016/j.crci.2015.10.002/},
	doi = {10.1016/j.crci.2015.10.002},
	language = {fr},
	number = {1-2},
	urldate = {2025-05-16},
	journal = {Comptes Rendus. Chimie},
	author = {Huix-Rotllant, Miquel and Burghardt, Irene and Ferré, Nicolas},
	year = {2016},
	pages = {50--56},
}

@book{gilbert_essentials_1991,
	address = {Boca Raton},
	title = {Essentials of {Molecular} {Photochemistry}},
	isbn = {978-0-8493-7727-3},
	language = {English},
	publisher = {CRC Press},
	author = {Gilbert, Andrew and Baggott, Jim},
	year = {1991},
}

@article{wolf_transient_2021,
	title = {Transient resonant {Auger}–{Meitner} spectra of photoexcited thymine},
	volume = {228},
	issn = {1364-5498},
	url = {https://pubs.rsc.org/en/content/articlelanding/2021/fd/d0fd00112k},
	doi = {10.1039/D0FD00112K},
	language = {en},
	number = {0},
	urldate = {2024-02-13},
	journal = {Faraday Discussions},
	author = {Wolf, Thomas J. A. and Paul, Alexander C. and Folkestad, Sarai D. and Myhre, Rolf H. and Cryan, James P. and Berrah, Nora and Bucksbaum, Phil H. and Coriani, Sonia and Coslovich, Giacomo and Feifel, Raimund and Martinez, Todd J. and Moeller, Stefan P. and Mucke, Melanie and Obaid, Razib and Plekan, Oksana and Squibb, Richard J. and Koch, Henrik and Gühr, Markus},
	month = may,
	year = {2021},
	note = {Publisher: The Royal Society of Chemistry},
	pages = {555--570},
}

@article{wang_imaging_2025,
	title = {Imaging the photochemical dynamics of cyclobutanone with {MeV} ultrafast electron diffraction},
	volume = {162},
	issn = {0021-9606},
	url = {https://doi.org/10.1063/5.0267186},
	doi = {10.1063/5.0267186},
	number = {18},
	urldate = {2025-07-11},
	journal = {The Journal of Chemical Physics},
	author = {Wang, Tianyu and Jiang, Hui and Jin, Cheng and Zou, Xiao and Zhu, Pengfei and Jiang, Tao and He, Feng and Xiang, Dao},
	month = may,
	year = {2025},
	pages = {184201},
}

@article{peng_photodissociation_2024,
	title = {The photodissociation dynamics and ultrafast electron diffraction image of cyclobutanone from the surface hopping dynamics simulation},
	volume = {160},
	issn = {0021-9606},
	url = {https://doi.org/10.1063/5.0203462},
	doi = {10.1063/5.0203462},
	number = {22},
	urldate = {2024-09-20},
	journal = {The Journal of Chemical Physics},
	author = {Peng, Jiawei and Liu, Hong and Lan, Zhenggang},
	month = jun,
	year = {2024},
	pages = {224305},
}

@article{vindel-zandbergen_non-adiabatic_2024,
	title = {Non-adiabatic dynamics of photoexcited cyclobutanone: {Predicting} structural measurements from trajectory surface hopping with {XMS}-{CASPT2} simulations},
	volume = {161},
	issn = {0021-9606},
	shorttitle = {Non-adiabatic dynamics of photoexcited cyclobutanone},
	url = {https://doi.org/10.1063/5.0203722},
	doi = {10.1063/5.0203722},
	number = {2},
	urldate = {2024-09-20},
	journal = {The Journal of Chemical Physics},
	author = {Vindel-Zandbergen, Patricia and González-Vázquez, Jesús},
	month = jul,
	year = {2024},
	pages = {024104},
}

@article{suchan_prediction_2024,
	title = {Prediction challenge: {First} principles simulation of the ultrafast electron diffraction spectrum of cyclobutanone},
	volume = {160},
	issn = {0021-9606},
	shorttitle = {Prediction challenge},
	url = {https://doi.org/10.1063/5.0198333},
	doi = {10.1063/5.0198333},
	number = {13},
	urldate = {2024-09-20},
	journal = {The Journal of Chemical Physics},
	author = {Suchan, Jiří and Liang, Fangchun and Durden, Andrew S. and Levine, Benjamin G.},
	month = apr,
	year = {2024},
	pages = {134310},
}

@article{mukherjee_prediction_2024,
	title = {Prediction {Challenge}: {Simulating} {Rydberg} photoexcited cyclobutanone with surface hopping dynamics based on different electronic structure methods},
	volume = {160},
	issn = {0021-9606},
	shorttitle = {Prediction {Challenge}},
	url = {https://doi.org/10.1063/5.0203636},
	doi = {10.1063/5.0203636},
	number = {15},
	urldate = {2024-09-20},
	journal = {The Journal of Chemical Physics},
	author = {Mukherjee, Saikat and Mattos, Rafael S. and Toldo, Josene M. and Lischka, Hans and Barbatti, Mario},
	month = apr,
	year = {2024},
	pages = {154306},
}

@article{miller_ultrafast_2024,
	title = {Ultrafast photochemistry and electron diffraction for cyclobutanone in the {S2} state: {Surface} hopping with time-dependent density functional theory},
	volume = {161},
	issn = {0021-9606},
	shorttitle = {Ultrafast photochemistry and electron diffraction for cyclobutanone in the {S2} state},
	url = {https://doi.org/10.1063/5.0203679},
	doi = {10.1063/5.0203679},
	number = {3},
	urldate = {2024-09-20},
	journal = {The Journal of Chemical Physics},
	author = {Miller, Ericka Roy and Hoehn, Sean J. and Kumar, Abhijith and Jiang, Dehua and Parker, Shane M.},
	month = jul,
	year = {2024},
	pages = {034105},
}

@article{miao_casscfmrci_2024,
	title = {A {CASSCF}/{MRCI} trajectory surface hopping simulation of the photochemical dynamics and the gas phase ultrafast electron diffraction patterns of cyclobutanone},
	volume = {160},
	issn = {0021-9606},
	url = {https://doi.org/10.1063/5.0197768},
	doi = {10.1063/5.0197768},
	number = {12},
	urldate = {2024-09-20},
	journal = {The Journal of Chemical Physics},
	author = {Miao, Xincheng and Diemer, Kira and Mitrić, Roland},
	month = mar,
	year = {2024},
	pages = {124309},
}

@article{makhov_ultrafast_2024,
	title = {Ultrafast electron diffraction of photoexcited gas-phase cyclobutanone predicted by ab initio multiple cloning simulations},
	volume = {160},
	issn = {0021-9606},
	url = {https://doi.org/10.1063/5.0203683},
	doi = {10.1063/5.0203683},
	number = {16},
	urldate = {2024-09-20},
	journal = {The Journal of Chemical Physics},
	author = {Makhov, Dmitry V. and Hutton, Lewis and Kirrander, Adam and Shalashilin, Dmitrii V.},
	month = apr,
	year = {2024},
	pages = {164310},
}

@article{janos_predicting_2024,
	title = {Predicting the photodynamics of cyclobutanone triggered by a laser pulse at 200 nm and its {MeV}-{UED} signals—{A} trajectory surface hopping and {XMS}-{CASPT2} perspective},
	volume = {160},
	issn = {0021-9606},
	url = {https://doi.org/10.1063/5.0203105},
	doi = {10.1063/5.0203105},
	number = {14},
	urldate = {2024-09-20},
	journal = {The Journal of Chemical Physics},
	author = {Janoš, Jiří and Figueira Nunes, Joao Pedro and Hollas, Daniel and Slavíček, Petr and Curchod, Basile F. E.},
	month = apr,
	year = {2024},
	pages = {144305},
}

@article{jaiswal_ultrafast_2024,
	title = {Ultrafast photochemistry and electron-diffraction spectra in n → (3s) {Rydberg} excited cyclobutanone resolved at the multireference perturbative level},
	volume = {160},
	issn = {0021-9606},
	url = {https://doi.org/10.1063/5.0203624},
	doi = {10.1063/5.0203624},
	number = {16},
	urldate = {2024-09-20},
	journal = {The Journal of Chemical Physics},
	author = {Jaiswal, V. K. and Montorsi, F. and Aleotti, F. and Segatta, F. and Keefer, Daniel and Mukamel, Shaul and Nenov, A. and Conti, I. and Garavelli, M.},
	month = apr,
	year = {2024},
	pages = {164316},
}

@article{hutton_using_2024,
	title = {Using a multistate mapping approach to surface hopping to predict the ultrafast electron diffraction signal of gas-phase cyclobutanone},
	volume = {160},
	issn = {0021-9606},
	url = {https://doi.org/10.1063/5.0203667},
	doi = {10.1063/5.0203667},
	number = {20},
	urldate = {2024-09-20},
	journal = {The Journal of Chemical Physics},
	author = {Hutton, Lewis and Moreno Carrascosa, Andrés and Prentice, Andrew W. and Simmermacher, Mats and Runeson, Johan E. and Paterson, Martin J. and Kirrander, Adam},
	month = may,
	year = {2024},
	pages = {204307},
}

@article{bennett_prediction_2024,
	title = {Prediction through quantum dynamics simulations: {Photo}-excited cyclobutanone},
	volume = {160},
	issn = {0021-9606},
	shorttitle = {Prediction through quantum dynamics simulations},
	url = {https://doi.org/10.1063/5.0203654},
	doi = {10.1063/5.0203654},
	number = {17},
	urldate = {2024-09-20},
	journal = {The Journal of Chemical Physics},
	author = {Bennett, Olivia and Freibert, Antonia and Spinlove, K. Eryn and Worth, Graham A.},
	month = may,
	year = {2024},
	pages = {174305},
}

@article{martin_santa_daria_photofragmentation_2024,
	title = {Photofragmentation of cyclobutanone at 200 nm: {TDDFT} vs {CASSCF} electron diffraction},
	volume = {160},
	issn = {0021-9606},
	shorttitle = {Photofragmentation of cyclobutanone at 200 nm},
	url = {https://doi.org/10.1063/5.0197895},
	doi = {10.1063/5.0197895},
	number = {11},
	urldate = {2024-09-20},
	journal = {The Journal of Chemical Physics},
	author = {Martín Santa Daría, Alberto and Hernández-Rodríguez, Javier and Ibele, Lea M. and Gómez, Sandra},
	month = mar,
	year = {2024},
	pages = {114303},
}

@article{lawrence_mash_2024,
	title = {A {MASH} simulation of the photoexcited dynamics of cyclobutanone},
	volume = {160},
	issn = {0021-9606},
	url = {https://doi.org/10.1063/5.0203695},
	doi = {10.1063/5.0203695},
	number = {17},
	urldate = {2024-09-20},
	journal = {The Journal of Chemical Physics},
	author = {Lawrence, Joseph E. and Ansari, Imaad M. and Mannouch, Jonathan R. and Manae, Meghna A. and Asnaashari, Kasra and Kelly, Aaron and Richardson, Jeremy O.},
	month = may,
	year = {2024},
	pages = {174306},
}

@article{hait_prediction_2024,
	title = {Prediction of photodynamics of 200 nm excited cyclobutanone with linear response electronic structure and ab initio multiple spawning},
	volume = {160},
	issn = {0021-9606},
	url = {https://doi.org/10.1063/5.0203800},
	doi = {10.1063/5.0203800},
	number = {24},
	urldate = {2024-09-20},
	journal = {The Journal of Chemical Physics},
	author = {Hait, Diptarka and Lahana, Dean and Fajen, O. Jonathan and Paz, Amiel S. P. and Unzueta, Pablo A. and Rana, Bhaskar and Lu, Lixin and Wang, Yuanheng and Kjønstad, Eirik F. and Koch, Henrik and Martínez, Todd J.},
	month = jun,
	year = {2024},
	pages = {244101},
}

@article{eng_photochemistry_2024,
	title = {The photochemistry of {Rydberg}-excited cyclobutanone: {Photoinduced} processes and ground state dynamics},
	volume = {160},
	issn = {0021-9606},
	shorttitle = {The photochemistry of {Rydberg}-excited cyclobutanone},
	url = {https://doi.org/10.1063/5.0203597},
	doi = {10.1063/5.0203597},
	number = {15},
	urldate = {2024-09-20},
	journal = {The Journal of Chemical Physics},
	author = {Eng, J. and Rankine, C. D. and Penfold, T. J.},
	month = apr,
	year = {2024},
	pages = {154301},
}

@article{green_imaging_2025,
	title = {Imaging the photochemistry of cyclobutanone using ultrafast electron diffraction: {Experimental} results},
	volume = {162},
	issn = {0021-9606},
	shorttitle = {Imaging the photochemistry of cyclobutanone using ultrafast electron diffraction},
	url = {https://doi.org/10.1063/5.0266559},
	doi = {10.1063/5.0266559},
	number = {18},
	urldate = {2025-05-12},
	journal = {The Journal of Chemical Physics},
	author = {Green, A. E. and Liu, Y. and Allum, F. and Graßl, M. and Lenzen, P. and Ashfold, M. N. R. and Bhattacharyya, S. and Cheng, X. and Centurion, M. and Crane, S. W. and Forbes, R. and Goff, N. A. and Huang, L. and Kaufman, B. and Kling, M.-F. and Kramer, P. L. and Lam, H. V. S. and Larsen, K. A. and Lemons, R. and Lin, M.-F. and Orr-Ewing, A. J. and Rolles, D. and Rudenko, A. and Saha, S. K. and Searles, J. and Shen, X. and Weathersby, S. and Weber, P. M. and Zhao, H. and Wolf, T. J. A.},
	month = may,
	year = {2025},
	pages = {184303},
}

@article{driver_attosecond_2024,
    title = {Attosecond delays in {X}-ray molecular ionization},
    volume = {632},
    copyright = {2024 The Author(s), under exclusive licence to Springer Nature Limited},
    issn = {1476-4687},
    url = {https://www.nature.com/articles/s41586-024-07771-9},
    doi = {10.1038/s41586-024-07771-9},
    language = {en},
    number = {8026},
    urldate = {2024-08-22},
    journal = {Nature},
    author = {Driver, Taran and Mountney, Miles and Wang, Jun and Ortmann, Lisa and Al-Haddad, Andre and Berrah, Nora and Bostedt, Christoph and Champenois, Elio G. and DiMauro, Louis F. and Duris, Joseph and Garratt, Douglas and Glownia, James M. and Guo, Zhaoheng and Haxton, Daniel and Isele, Erik and Ivanov, Igor and Ji, Jiabao and Kamalov, Andrei and Li, Siqi and Lin, Ming-Fu and Marangos, Jon P. and Obaid, Razib and O’Neal, Jordan T. and Rosenberger, Philipp and Shivaram, Niranjan H. and Wang, Anna L. and Walter, Peter and Wolf, Thomas J. A. and Wörner, Hans Jakob and Zhang, Zhen and Bucksbaum, Philip H. and Kling, Matthias F. and Landsman, Alexandra S. and Lucchese, Robert R. and Emmanouilidou, Agapi and Marinelli, Agostino and Cryan, James P.},
    month = aug,
    year = {2024},
    note = {Publisher: Nature Publishing Group},
    pages = {762--767},
}

@article{stellato_osqp_2020,
	title = {{OSQP}: an operator splitting solver for quadratic programs},
	volume = {12},
	issn = {1867-2957},
	shorttitle = {{OSQP}},
	url = {https://doi.org/10.1007/s12532-020-00179-2},
	doi = {10.1007/s12532-020-00179-2},
	language = {en},
	number = {4},
	urldate = {2025-06-19},
	journal = {Mathematical Programming Computation},
	author = {Stellato, Bartolomeo and Banjac, Goran and Goulart, Paul and Bemporad, Alberto and Boyd, Stephen},
	month = dec,
	year = {2020},
	pages = {637--672},
}

@article{wang_photon_2023,
	title = {Photon energy-resolved velocity map imaging from spectral domain ghost imaging},
	volume = {25},
	issn = {1367-2630},
	url = {https://dx.doi.org/10.1088/1367-2630/acc201},
	doi = {10.1088/1367-2630/acc201},
	language = {en},
	number = {3},
	urldate = {2025-06-19},
	journal = {New Journal of Physics},
	author = {Wang, Jun and Driver, Taran and Allum, Felix and Papadopoulou, Christina C and Passow, Christopher and Brenner, Günter and Li, Siqi and Düsterer, Stefan and Tul Noor, Atia and Kumar, Sonu and Bucksbaum, Philip H and Erk, Benjamin and Forbes, Ruaridh and Cryan, James P},
	month = mar,
	year = {2023},
	note = {Publisher: IOP Publishing},
	pages = {033017},
}

@article{hohenstein_predictions_2021,
	title = {Predictions of {Pre}-edge {Features} in {Time}-{Resolved} {Near}-{Edge} {X}-ray {Absorption} {Fine} {Structure} {Spectroscopy} from {Hole}–{Hole} {Tamm}–{Dancoff}-{Approximated} {Density} {Functional} {Theory}},
	volume = {17},
	issn = {1549-9618},
	url = {https://doi.org/10.1021/acs.jctc.1c00478},
	doi = {10.1021/acs.jctc.1c00478},
	number = {11},
	urldate = {2025-09-25},
	journal = {Journal of Chemical Theory and Computation},
	author = {Hohenstein, Edward G. and Yu, Jimmy K. and Bannwarth, Christoph and List, Nanna Holmgaard and Paul, Alexander C. and Folkestad, Sarai D. and Koch, Henrik and Martínez, Todd J.},
	month = nov,
	year = {2021},
	note = {Publisher: American Chemical Society},
	pages = {7120--7133},
}

@article{levine_implementation_2008,
	series = {Ultrafast {Photoinduced} {Processes} in {Polyatomic} {Molecules}},
	title = {Implementation of \textit{ab initio} multiple spawning in the {Molpro} quantum chemistry package},
	volume = {347},
	issn = {0301-0104},
	url = {https://www.sciencedirect.com/science/article/pii/S0301010408000190},
	doi = {10.1016/j.chemphys.2008.01.014},
	number = {1},
	urldate = {2025-09-29},
	journal = {Chemical Physics},
	author = {Levine, Benjamin G. and Coe, Joshua D. and Virshup, Aaron M. and Martínez, Todd J.},
	month = may,
	year = {2008},
	pages = {3--16},
}

@article{facciala_unraveling_2025,
    title = {Unraveling the {Relaxation} {Dynamics} of {Uracil}: {Insights} from {Time}-{Resolved} {X}-ray {Photoelectron} {Spectroscopy}},
    volume = {147},
    issn = {0002-7863},
    shorttitle = {Unraveling the {Relaxation} {Dynamics} of {Uracil}},
    url = {https://doi.org/10.1021/jacs.5c04874},
    doi = {10.1021/jacs.5c04874},
    number = {34},
    urldate = {2025-10-20},
    journal = {Journal of the American Chemical Society},
    publisher = {American Chemical Society},
    author = {Faccialà, Davide and Bonanomi, Matteo and Tenorio, Bruno Nunes Cabral and Avaldi, Lorenzo and Bolognesi, Paola and Callegari, Carlo and Coreno, Marcello and Coriani, Sonia and Decleva, Piero and Devetta, Michele and Došlić, Nađa and De Fanis, Alberto and Di Fraia, Michele and Lever, Fabiano and Mazza, Tommaso and Meyer, Michael and Mullins, Terry and Ovcharenko, Yevheniy and Pal, Nitish and Piancastelli, Maria Novella and Richter, Robert and Rivas, Daniel E. and Sapunar, Marin and Senfftleben, Björn and Usenko, Sergey and Vozzi, Caterina and Gühr, Markus and Prince, Kevin C. and Plekan, Oksana},
    month = aug,
    year = {2025},
    pages = {30694--30707},
}

@article{mayer_following_2022,
    title = {Following excited-state chemical shifts in molecular ultrafast x-ray photoelectron spectroscopy},
    volume = {13},
    copyright = {2022 The Author(s)},
    issn = {2041-1723},
    url = {https://www.nature.com/articles/s41467-021-27908-y},
    doi = {10.1038/s41467-021-27908-y},
    language = {en},
    number = {1},
    urldate = {2022-05-09},
    journal = {Nature Communications},
    publisher = {Nature Publishing Group},
    author = {Mayer, D. and Lever, F. and Picconi, D. and Metje, J. and Alisauskas, S. and Calegari, F. and Düsterer, S. and Ehlert, C. and Feifel, R. and Niebuhr, M. and Manschwetus, B. and Kuhlmann, M. and Mazza, T. and Robinson, M. S. and Squibb, R. J. and Trabattoni, A. and Wallner, M. and Saalfrank, P. and Wolf, T. J. A. and Gühr, M.},
    month = jan,
    year = {2022},
    note = {Number: 1},
    pages = {198},
}

@article{stankus_ultrafast_2019,
    title = {Ultrafast {X}-ray scattering reveals vibrational coherence following {Rydberg} excitation},
    volume = {11},
    issn = {1755-4330, 1755-4349},
    url = {http://www.nature.com/articles/s41557-019-0291-0},
    doi = {10.1038/s41557-019-0291-0},
    language = {en},
    urldate = {2019-07-15},
    journal = {Nature Chemistry},
    author = {Stankus, Brian and Yong, Haiwang and Zotev, Nikola and Ruddock, Jennifer M. and Bellshaw, Darren and Lane, Thomas J. and Liang, Mengning and Boutet, Sébastien and Carbajo, Sergio and Robinson, Joseph S. and Du, Wenpeng and Goff, Nathan and Chang, Yu and Koglin, Jason E. and Minitti, Michael P. and Kirrander, Adam and Weber, Peter M.},
    month = jul,
    year = {2019},
    pages = {716 -- 721},
}

@article{wolf_photochemical_2019,
	title = {The photochemical ring-opening of 1,3-cyclohexadiene imaged by ultrafast electron diffraction},
	volume = {11},
	copyright = {2019 This is a U.S. government work and not under copyright protection in the U.S.; foreign copyright protection may apply},
	issn = {1755-4349},
	url = {http://www.nature.com/articles/s41557-019-0252-7},
	doi = {10.1038/s41557-019-0252-7},
	language = {En},
	number = {6},
	urldate = {2019-04-15},
	journal = {Nature Chemistry},
	author = {Wolf, T. J. A. and Sanchez, D. M. and Yang, J. and Parrish, R. M. and Nunes, J. P. F. and Centurion, M. and Coffee, R. and Cryan, J. P. and Gühr, M. and Hegazy, K. and Kirrander, A. and Li, R. K. and Ruddock, J. and Shen, X. and Vecchione, T. and Weathersby, S. P. and Weber, P. M. and Wilkin, K. and Yong, H. and Zheng, Q. and Wang, X. J. and Minitti, M. P. and Martínez, T. J.},
	month = apr,
	year = {2019},
	pages = {504 -- 509},
}

@article{blanchet_discerning_1999,
    title = {Discerning vibronic molecular dynamics using time-resolved photoelectron spectroscopy},
    volume = {401},
    issn = {0028-0836},
    url = {http://dx.doi.org/10.1038/43410},
    number = {6748},
    journal = {Nature},
    author = {Blanchet, Valerie and Zgierski, Marek Z. and Seideman, Tamar and Stolow, Albert},
    month = sep,
    year = {1999},
    pages = {52--54},
}

@article{attar_femtosecond_2017,
    title = {Femtosecond x-ray spectroscopy of an electrocyclic ring-opening reaction},
    volume = {356},
    copyright = {Copyright © 2017, American Association for the Advancement of Science},
    issn = {0036-8075, 1095-9203},
    url = {http://science.sciencemag.org/content/356/6333/54},
    doi = {10.1126/science.aaj2198},
    language = {en},
    number = {6333},
    urldate = {2017-04-08},
    journal = {Science},
    author = {Attar, Andrew R. and Bhattacherjee, Aditi and Pemmaraju, C. D. and Schnorr, Kirsten and Closser, Kristina D. and Prendergast, David and Leone, Stephen R.},
    month = apr,
    year = {2017},
    pages = {54--59},
}

@article{pertot_time-resolved_2017,
    title = {Time-resolved x-ray absorption spectroscopy with a water window high-harmonic source},
    volume = {355},
    copyright = {Copyright © 2017, American Association for the Advancement of Science},
    issn = {0036-8075, 1095-9203},
    url = {http://science.sciencemag.org/content/355/6322/264},
    doi = {10.1126/science.aah6114},
    language = {en},
    number = {6322},
    urldate = {2017-01-24},
    journal = {Science},
    author = {Pertot, Yoann and Schmidt, Cédric and Matthews, Mary and Chauvet, Adrien and Huppert, Martin and Svoboda, Vit and Conta, Aaron von and Tehlar, Andres and Baykusheva, Denitsa and Wolf, Jean-Pierre and Wörner, Hans Jakob},
    month = jan,
    year = {2017},
    pages = {264--267},
}

@article{chakraborty_time-resolved_2024,
    title = {Time-resolved photoelectron spectroscopy via trajectory surface hopping},
    volume = {14},
    copyright = {© 2024 Wiley Periodicals LLC.},
    issn = {1759-0884},
    url = {https://onlinelibrary.wiley.com/doi/abs/10.1002/wcms.1715},
    doi = {10.1002/wcms.1715},
    language = {en},
    number = {3},
    urldate = {2024-05-19},
    journal = {WIREs Computational Molecular Science},
    author = {Chakraborty, Pratip and Matsika, Spiridoula},
    year = {2024},
    note = {\_eprint: https://onlinelibrary.wiley.com/doi/pdf/10.1002/wcms.1715},
    pages = {e1715},
}

@article{li_imaging_2025,
	title = {Imaging a light-induced molecular elimination reaction with an {X}-ray free-electron laser},
	volume = {16},
	copyright = {2025 The Author(s)},
	issn = {2041-1723},
	url = {https://www.nature.com/articles/s41467-025-62274-z},
	doi = {10.1038/s41467-025-62274-z},
	language = {en},
	number = {1},
	urldate = {2026-03-08},
	journal = {Nature Communications},
	publisher = {Nature Publishing Group},
	author = {Li, Xiang and Boll, Rebecca and Vindel-Zandbergen, Patricia and González-Vázquez, Jesús and Rivas, Daniel E. and Bhattacharyya, Surjendu and Borne, Kurtis and Chen, Keyu and De Fanis, Alberto and Erk, Benjamin and Forbes, Ruaridh and Green, Alice E. and Ilchen, Markus and Kaderiya, Balram and Kukk, Edwin and Lam, Huynh Van Sa and Mazza, Tommaso and Mullins, Terence and Senfftleben, Björn and Trinter, Florian and Usenko, Sergey and Venkatachalam, Anbu Selvam and Wang, Enliang and Cryan, James P. and Meyer, Michael and Jahnke, Till and Ho, Phay J. and Rolles, Daniel and Rudenko, Artem},
	month = jul,
	year = {2025},
	pages = {7006},
}
\end{document}


\maketitle

\section{Spectral domain ghost imaging} \label{SI:spooktroscopy}
\begin{figure}[!ht]
    \centering
    \includegraphics[width=\textwidth]{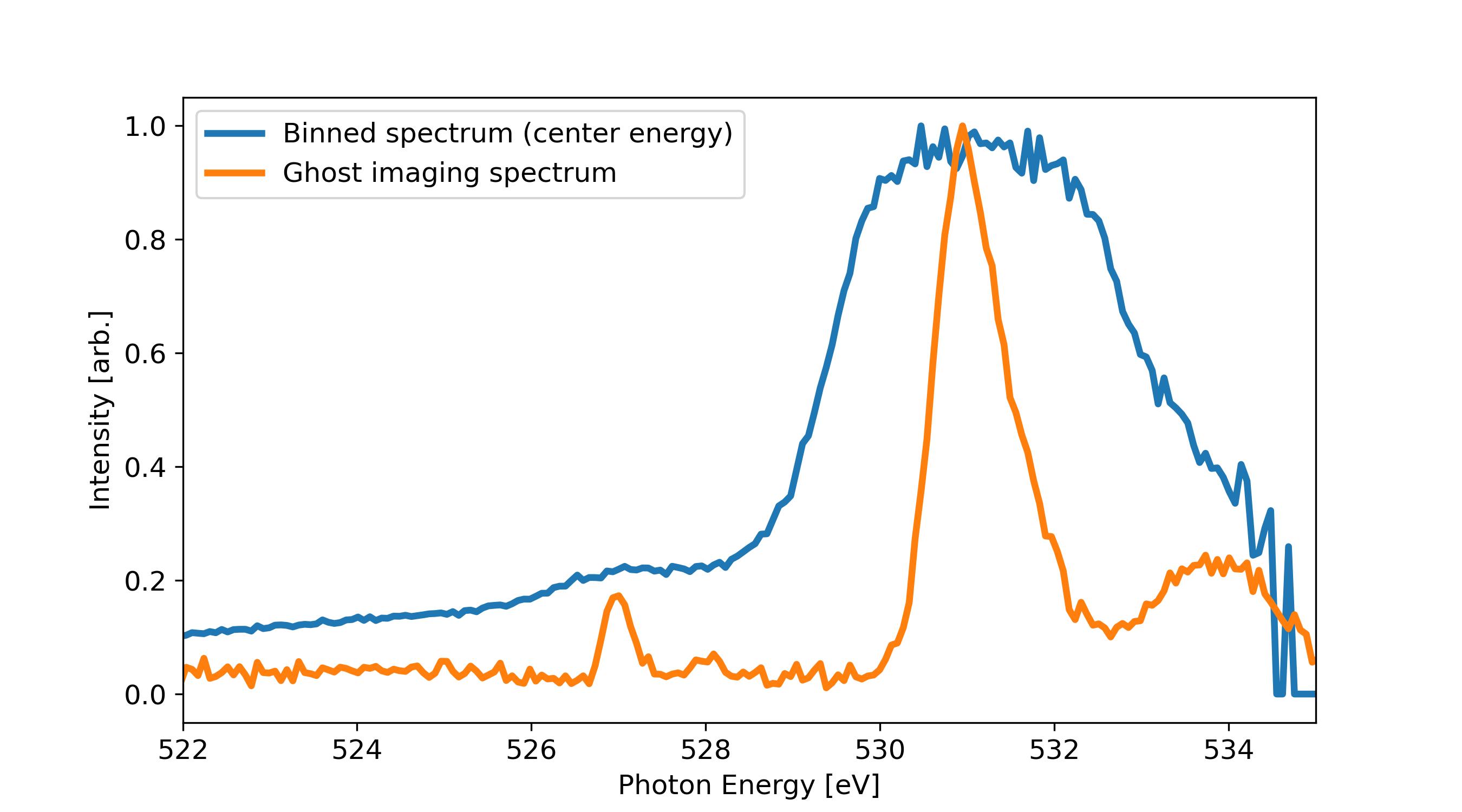}
    \caption{Comparison spectral domain ghost imaging and central photon energy binning}
    \label{fig:spookvsCOM}
\end{figure}

To enhance the spectral resolution of the measurement, we employed the ghost imaging technique, which has been successfully implemented in various FEL experiments \cite{li_two-dimensional_2021,wang_photon_2023, li_ghost-imaging-enhanced_2022}. Thereby the absorption spectrum is not directly measured, but reconstructed by correlating the spectrum of the incident light source with a 'bucket' measurement of the total absorption in the sample on a shot-to-shot basis. In our case the total absorption is obtained by measuring the absolute yield of Auger-Meitner electrons originating from the decay of the core excited state. The photon spectrum is measured downstream of the interaction point, after interaction with the sample. However, the low density of the sample gas leads to negligible attenuation of the X-rays. Therefore, the transmitted spectrum is essentially identical to the incident spectrum. The ghost imaging problem can be expressed in a compact mathematical form using a simple linear matrix multiplication:

\begin{equation}
    \textbf{b} = \textbf{A}\textbf{x}
    \label{eq:spooktroskopy}
\end{equation}

with a vector $\textbf{b}$, containing the bucket measurements of the total electron yield for $n$ measurements and $\textbf{A}$, the $n\times m$ matrix containing the corresponding spectra of the incident light measured with a resolution of $m$ pixels. The vector $\textbf{x}$ denotes the unknown spectral response of the system, which is the quantity to be reconstructed. Equation \ref{eq:spooktroskopy} can be solved using the pseudo-inverse of $\textbf{A}$ \cite{wang_photon_2023,stellato_osqp_2020}. This method is sensitive to noise, and regularizations of x are typically applied to mitigate its effects. Generally the absorption of the sample can be assumed to be smooth, sparse, and nonnegative. In our case the data quality is good enough to ignore smoothness and sparsity and only restrict our solution by forcing a positive response. (Note: The negative bleach signature in figure 3 in the main text is caused by calculating the difference spectrum after using ghost imaging on the pumped and unpumped data. The respective absorption spectra are still all positive, see figure 2 in the main text). While ignoring smoothness and sparsity regularizations increases the complexity of the reconstruction and results in a higher noise level, it also mitigates the risk of information loss due to over-regularization. By applying this technique, the spectral resolution of the measurement becomes limited solely by the resolution of the spectrometer and the intrinsic variation of the FEL, rather than by the bandwidth of the incident light source. Figure \ref{fig:spookvsCOM} shows a comparison between the data obtained by spectral domain ghost imaging and data binning using the central photon energy of the incident light. The conventional binning approach, shown in blue, exhibits a broad single peak centered at 531 eV with a weak shoulder near 527 eV. In contrast, application of the ghost imaging method, shown in orange, resolves the shoulder into a distinct peak and reveals that the main feature comprises a double-peak structure.

An important prerequisite for accurate spectral reconstruction using ghost imaging is a linear detector response, since the method relies on solving a linear inverse problem. Nonlinearities due to detector saturation lead to distortions in the reconstructed absorption spectrum. For the experiment, the magnetic bottle spectrometer was optimized to resolve the transient feature at 527 eV (see Figure \ref{fig:spookvsCOM}), which is significantly weaker than the static ground-state feature at 531 eV. This optimization resulted in substantial detector saturation when scanning across the ground-state resonance. To mitigate the impact of saturation on the reconstruction of the transient signal, we restricted the dataset to shots exhibiting only minimal spectral intensity near the ground-state resonance. However, due to the presence of saturated shots and reduced statistics in the vicinity of the ground-state feature, the resulting data exhibit a lower signal-to-noise ratio, and the reliability of the extracted signal in this region is correspondingly limited.

\section{Acetophenone ground state absorption spectrum} \label{SI:groundstate}

\begin{figure}[!ht]
    \centering
    \includegraphics[width=\textwidth]{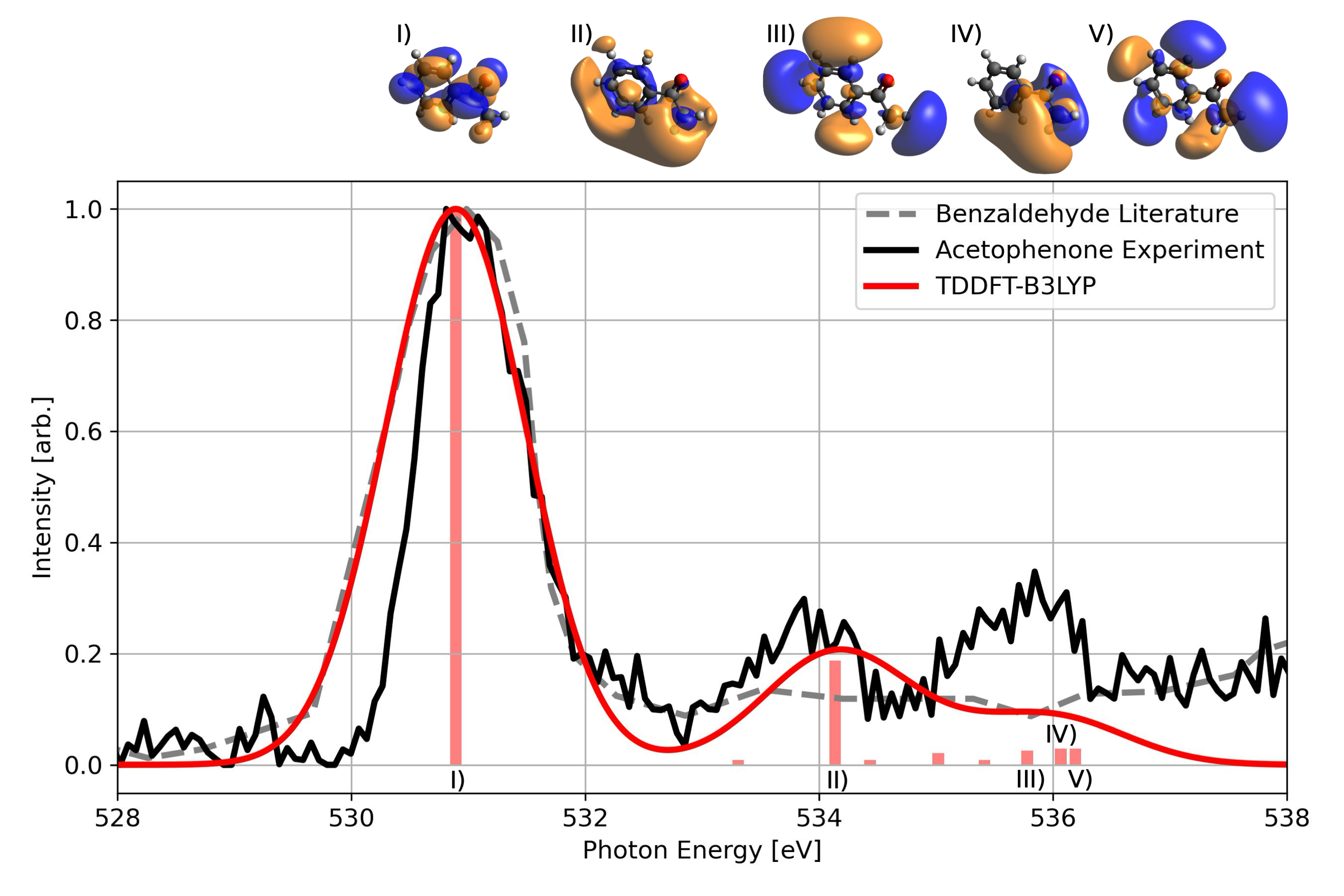}
    \caption{Comparison of experimental (black) and calculated (red) AP ground state absorption spectrum. The calculation was performed at the TD-DFT/B3LYP/def2-QZYPP level, the relevant orbitals with the hightest oscillatory strength are shown above the plot. The spectrum of the structurally similar benzaldehyde from the literature \cite{hitchcock_inner-shell_1992} (drashed grey) is given as a reference for the the main excitation peak. In addition to the main absorption peak at 531~eV we are able to observe two peaks at 533.9~eV and 535.7~eV that are reasonably well predicted by the calculations.}
    \label{fig:absop_theory}
\end{figure}

To the best of our knowledge, oxygen K-edge NEXAFS spectra for acetophenone have not been previously reported in the literature. To address this knowledge gap, we measured the static ground-state absorption spectrum of the sample over an extended photon energy range to obtain a comprehensive view of the transitions leading up to the K-edge ionization. The spectrum is plotted in figure \ref{fig:absop_theory} (black), it exhibits a main abortion peak at 531~eV followed by two weaker peaks at 533.9 eV and 535.7~eV. These peaks are reasonably reproduced by calculations performed at the TD-DFT/B3LYP/def2-QZYPP level using the quantum chemistry program ORCA \cite{neese_software_2022, neese_efficient_2002, neese_improvement_2003, neese_efficient_2009, helmich-paris_improved_2021, neese_shark_2023}. These calculation were performed independent of the calculations in the main text. The orbitals of the highest intensity contribution are shown above the plot. The corresponding energy values and oscillator strengths are given in table \ref{tab:theory_oscs}. Both the experimental and calculated spectrum are referenced against the structurally similar benzaldehyde taken from Hitchcock et al. \cite{hitchcock_inner-shell_1992}. To validate this comparison, we performed a TD-DFT/B3LYP/def2-QZYPP calculation for benzaldehyde and found the main absorption peak to be different in photon energy by only 0.04~eV. This shows that the substitution of one hydrogen atom with a methyl group at the carbonyl carbon does not affect the resonant energies at the oxygen 1s site. The absorption spectrum is determined by convoluting the calculated core excitation energies and associated oscillator strengths using Gaussian broadening with a FWHM of 1.4~eV. The spectrum is scaled in intensity to the main peak of the experimental spectrum and shifted in photon energy by 13.9~eV to match its spectral position. This shift is also applied to all energies in table \ref{tab:theory_oscs}. In addition to this theory reference we calibrated the Fresnel zone plate spectrometer using the absorption of CO$_{2}$ gas at the oxygen edge. 

The calculations successfully reproduce the second peak at 534.1~eV and indicate that the third peak does not arise from a single electronic transition, but rather from the superposition of three weaker transitions at 535.9, 536.1, and 536.2~eV. While the calculations include minor contributions from additional weak transitions, they do not account for the onset of the ionization edge beyond 536~eV. The resulting rise in the baseline leads to an artificially enhanced prominence of the peak in the experimental spectra. The elevated baseline is evident in our AP data as well as in the literature reference for benzaldehyde.

\begin{table}[htbp]
\centering
\caption{Absorption energies and oscillator strengths for the TD-DFT/B3LYP calculation in figure \ref{fig:absop_theory}. The energies are shifted by 13.9~eV to match the main absorption peak of the measured spectrum at 530.9~eV. }
\begin{tabular}{cc}
\hline
Energy (eV) & f$_{\text{osc}}$ \\
\hline
530.893475 & 0.030445967 \\
533.308195 & 0.000195632 \\
534.136214 & 0.005767118 \\
534.434378 & 0.000251398 \\
535.017641 & 0.000699203 \\
535.413308 & 0.000241454 \\
535.780183 & 0.000829874 \\
535.899530 & 0.000009929 \\
536.066134 & 0.000848473 \\
536.190380 & 0.000860542 \\
\hline
\end{tabular}
\label{tab:theory_oscs}
\end{table}

\section{UV Power scan} \label{SI:UVpower}

\begin{figure}[!ht]
    \centering
    \includegraphics[width=\textwidth]{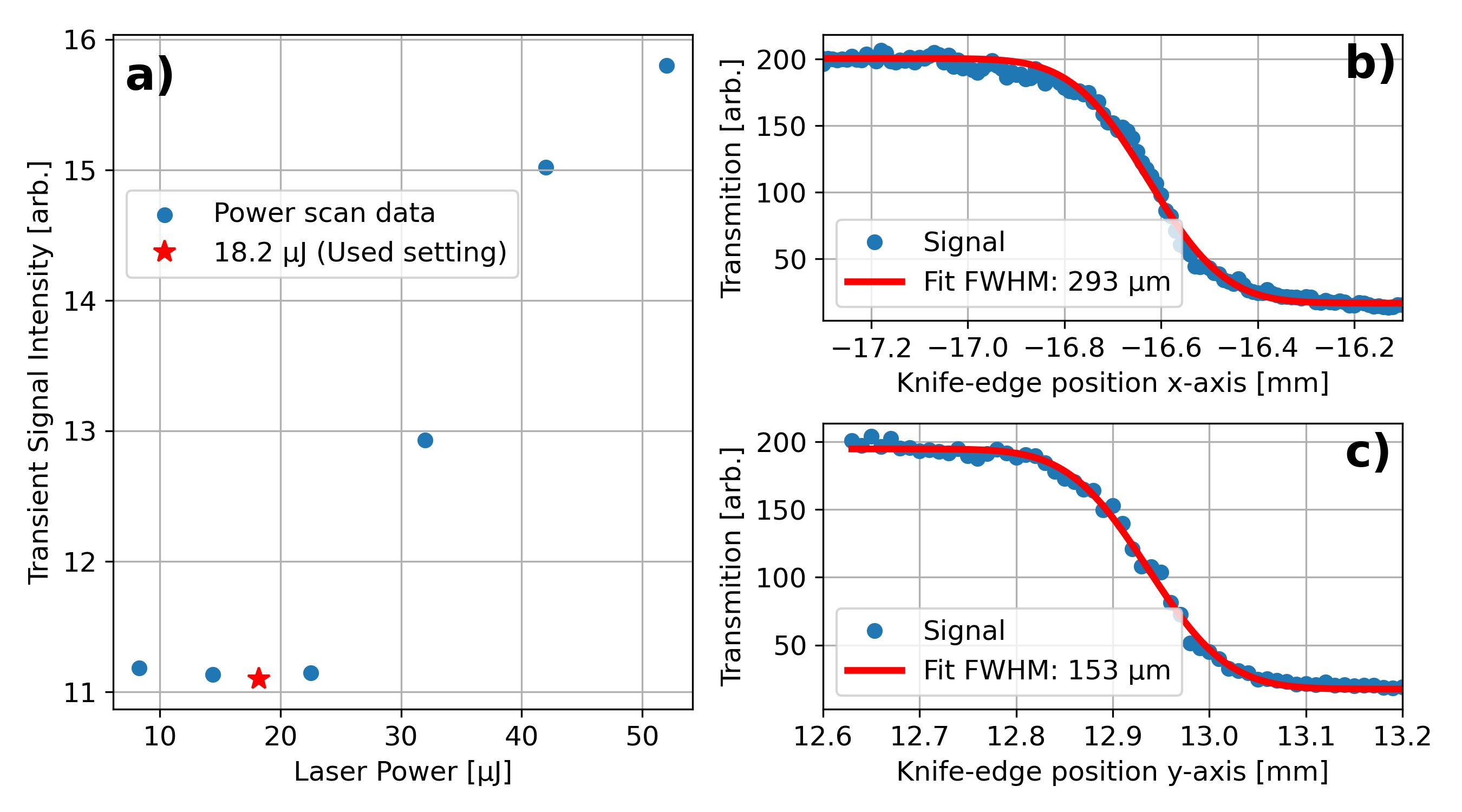}
    \caption{UV Laser characterization. a) The intensity of the transient signal (blue dot) shows no change for laser powers smaller than 25 $\mu$J and a close-to-linear dependence at higher pulse enegies. The laser pulse energy used during the experiment of 18.2 $\mu$J (red star) lies within a region where the signal-to-noise level was insufficient to show a linear dependence. b) and c) knife edge scan of the laser spotsize at the focus for the x- and y-axis respectively. The measured intensity of the transmitted signal (blue dots) is fitted with an error function (red line). The FWHM of the laser focus is 293 $\mu$m in x direction and 153 $\mu$m in y direction.}
    \label{fig:KnifeEdge}
\end{figure}

A 266~nm laser pulse with a pulse duration of 45~fs was used to excite the AP sample. To ensure that the observed transient features are exclusively resulting from single photon excitation, we measured the intensity of the UV induced feature at 527~eV at a delay of 2~ps for different laser powers. The absorption spectrum for each energy was reconstructed using the ghost imaging approach. The transient signal intensity corresponds to the integrated area of the 527~eV peak in the differential absorption spectrum obtained by subtracting the unpumped spectrum from the pumped spectrum. The results of the power scan are shown in figure \ref{fig:KnifeEdge} a) where the intensity of the transient signal is plotted against the UV pulse energy (blue dots). The red star marks the UV pulse energy of 18.2~$\mu$J that was used during the experiment. We did not observe a significant change in signal intensity for pulse energies lower than 25~$\mu$J. This plateau is caused by the ghost imaging method that struggles to reconstruct the spectrum with weak signal intensity in combination with a low signal to noise ratio. However, we observe an essentially linear dependence of the transient signal intensity on the UV intensity for significantly higher intensities than employed in the experiment. To further verify that the sample is not excessively pumped we measured the focus spot size using a knife edge scan figure (\ref{fig:KnifeEdge} b) and c)). The FWHM of the focus spot was measured to be 293~$\mu$m and 153~$\mu$m in horizontal and vertical direction, respectively. With a average focus spot size of 212~$\mu$m the laser fluence during the experiment was 0.1034~J/cm$^2$. Acetophenone has an absorption cross-section of 2~Mbarn at 266~nm \cite{berger_photochemical_1975}. In comparison, similar experiments using a sample with an absorption cross-section of 30~Mbarn and a pump fluence of 0.2546~J/cm² operated within the linear absorption regime \cite{wolf_probing_2017}.  

\section{Kinetic Model of the Population Dynamics}
\label{SI:kinModel}

The population dynamics across the excited states of AP are quantified using a kinetic model of 2 consecutive first-order processes,




\begin{equation}
A_1 \xrightarrow{k_1} A_2 \xrightarrow{k_2} A_3 ,
\end{equation}

where $A_1$ denotes the initially excited state, $A_2$ an intermediate state, and $A_3$ the final state. The corresponding rate constants are $k_1$ and $k_2$, respectively. The time-dependent populations of these states are given by

\begin{subequations}\label{eq:4:concentrationEquations}
\begin{align}
    [A_1](t) &= [A_1]_0\, e^{-k_1 t}, \\
    [A_2](t) &= [A_1]_0\frac{k_1}{k_2 - k_1}\left( e^{-k_1 t} - e^{-k_2 t} \right), \\
    [A_3](t) &= [A_1]_0\left[ 1 - \frac{k_2}{k_2 - k_1} e^{-k_1 t}+ \frac{k_1}{k_2 - k_1} e^{-k_2 t} \right],
\end{align}
\end{subequations}

where $[A_1]_0$ is the initial population of the photoexcited state.
Equations~\ref{eq:4:concentrationEquations} describe the absolute populations of the involved states. 
The experimentally measured NEXAFS intensities do
not directly correspond to the population of a given state, as they are modulated by the absorption cross section of a given state.

Differences in absorption cross sections prevent a direct determination of absolute state populations, but they do not affect the extracted time constants. At the same time, these differences can be exploited to distinguish overlapping electronic states. In the present experiment, this fact is used to track population transfer from the $ ^1n\pi^* $ state to the $ ^3n\pi^* $ state. Although these states are close in energy, they exhibit different relative absorption cross sections, allowing their contributions to be disentangled.

To extract the relevant time constants, the population equations are adapted to the experimentally accessible observable. The measured signal is proportional to the population of the corresponding state, modified by its absorption cross section. The measured intensity can therefore be expressed by 

\begin{equation}
I_n(t) = \sigma_n[A_n](t) 
\end{equation}

where $ \sigma_n $ is the absorption cross section of the states $A_n$. The measured signal is assumed to originate from states that are not directly populated by the initial excitation and to be independent of contributions from other states. Under these assumptions, the total measured intensity is given by


\begin{align}
I(t) &= I_2(t) + I_3(t) \\
I(t) &= \sigma_2(\sigma_1[A_1]_0)\frac{k_1}{k_2 - k_1}
       \left( e^{-k_1 t} - e^{-k_2 t} \right)
       + \notag\\
     &\quad \sigma_3(\sigma_1[A_1]_0)\left[
       1 - \frac{k_2}{k_2 - k_1} e^{-k_1 t}
       + \frac{k_1}{k_2 - k_1} e^{-k_2 t}
       \right]
\end{align}

The total initial population $[A_1]_0$ is unknown. By defining the relative maximum measurable population amplitude as $I_0=\sigma_2(\sigma_1[A_1]_0)$ the signal is effectively normalized to the contribution of the $A_2$ state. Introducing the relative difference in absorption cross section between the state $A_3$ and $A_2$ as $\eta=\sigma_3/\sigma_2$ and using the relation between rate constant and characteristic decay time, $\tau = 1/k$, yields

\begin{equation}
I(t) =
I_0\,\frac{\tau_2}{\tau_2 - \tau_1}
\left( e^{-t/\tau_2} - e^{-t/\tau_1} \right)
+ \eta I_0 \left(
1 - \frac{\tau_2}{\tau_2 - \tau_1} e^{-t/\tau_2}
+ \frac{\tau_1}{\tau_2 - \tau_1} e^{-t/\tau_1}
\right).
\end{equation}

This expression is used as the model function to fit the experimental data and extract the characteristic time constants of the population transfer.

\section{Computational Details} \label{SI:Theory}

To simulate the static and transient x-ray absorption of acetophenone at the oxygen K-edge from first principles, we initially perform ground-state geometry optimization and a subsequent harmonic frequency calculation at the B3LYP-D3/6-31G* level of theory\cite{becke1993density,lee1988development,grimme2010consistent,ditchfield1971self}.  

\begin{figure}[!ht]
    \centering
    \includegraphics[width=\textwidth]{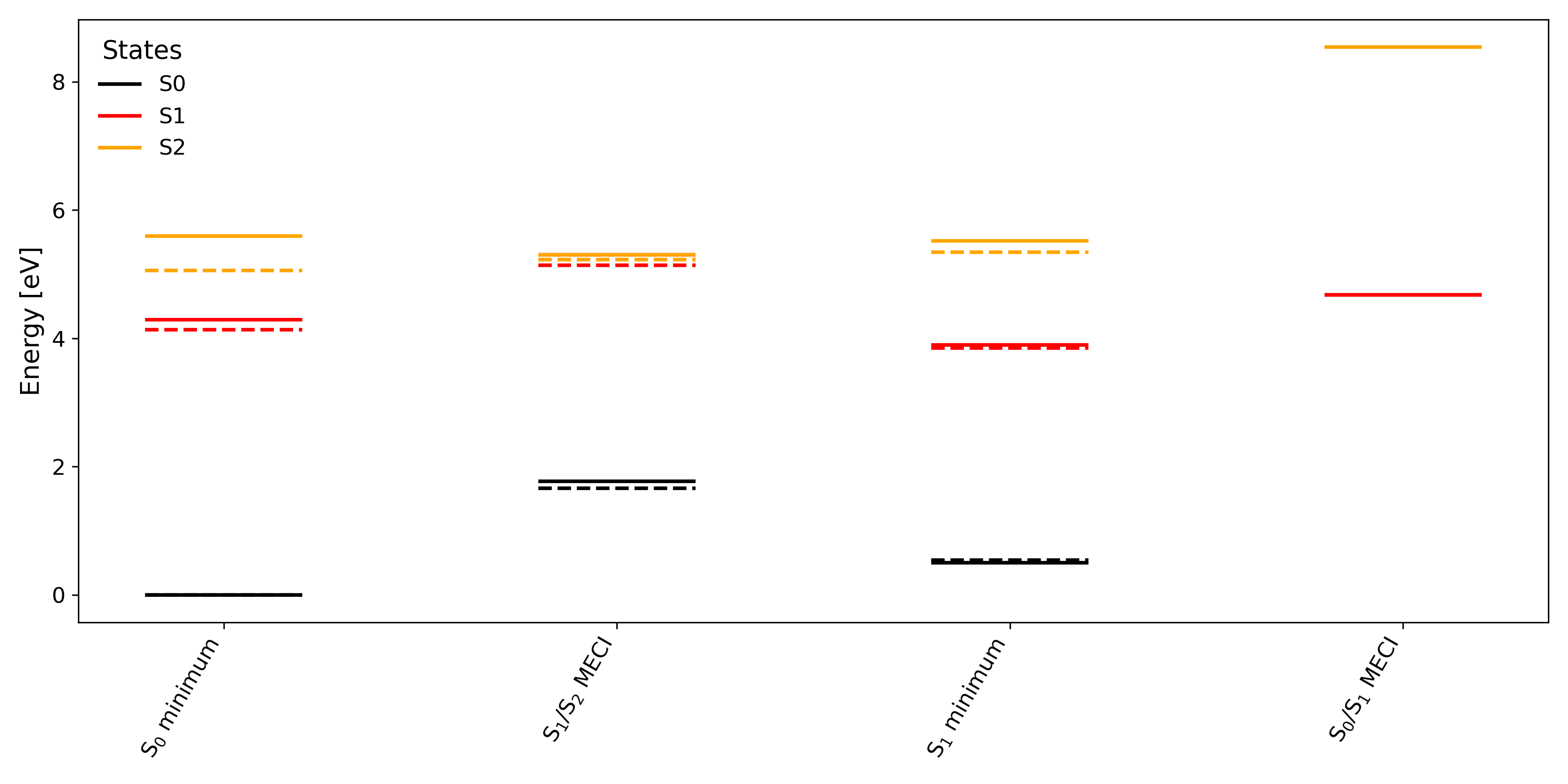}
    \caption{Energy diagram across different important geometries using fomo-hh-TDA-BHandHLYP (solid lines) and EOM-CCSD (dashed lines). Note: EOM-CCSD results are not shown for then S1/S0 minimum energy conical intersection (MECI).}
    \label{fig:energy_diagram}
\end{figure}

Following this, we conduct a series of calculations modeling the electronic structure of the system i.e., energies, gradients, nonadiabatic coupling, and transition moments, using the fomo-hh-TDA-BHandHLYP/def2-SVP methodology\cite{bannwarth_holehole_2020,yu2020ab,weigend2005balanced,becke1988density,lee1988development,becke1993new} to simulate x-ray absorption spectra. We chose fomo-hh-TDA-BHandHLYP based on its qualitative agreement with EOM-CCSD across different geometries as illustrated in Figure \ref{fig:energy_diagram} for all but the S1/S0 minimum energy conical intersection that EOM-CCSD is unable to describe meaningfully. For these calculations, we sample 50 random geometries around the ground-state minimum from a 0 K Wigner distribution\cite{wigner1932quantum} and determine vertical excitation energies and associated dipole oscillator strengths within the core-valence separation approximation\cite{cederbaum1980many} keeping only the oxygen 1s orbital active. In all cases, we model the x-ray absorption of the S$_0$, S$_1$, and S$_2$ states at the oxygen K-edge by calculating the excitation energies of the S$_1$ and S$_2$ states and of the nine lowest core-excited states from the oxygen 1s orbital as well as the oscillator strengths for transitions between all of these states. 

The static x-ray absorption spectra are determined by convoluting the vertical core excitation energies and associated oscillator strengths of each Wigner sample using Gaussian broadening with a full-width half maximum (FWHM) of 0.2~eV. Subsequently, the final spectra are obtained by averaging over the spectra obtained from the 50 geometries. 

To simulate excited state dynamics and transient x-ray absorption spectrum of acetophenone, we perform nonadiabatic molecular dynamics using the \textit{ab initio} multiple spawning (AIMS) method\cite{ben1998nonadiabatic,ben-nun_ab_2000}. We used the 50 Wigner samples (positions and momenta) as initial conditions and propagated the AIMS trajectory basis functions using a time step of 0.5~fs except in regions of high nonadiabatic coupling, where the step size is reduced to 0.25~fs to allow for more accurate integration of the time-derivative coupling. Following excitation to the bright S$_2$ ($\pi \rightarrow \pi^*$) state, we propagate each trajectory for 1 ps including the S$_0$, S$_1$, and S$_2$ states allowing for population transfer between these. Following the AIMS dynamics, we model the transient x-ray absorption spectrum by determining the x-ray absorption spectrum of each trajectory and averaging over them at each time step. For each trajectory, the contribution to the overall transient x-ray absorption spectrum is determined by computing the x-ray absorption through convolution using Gaussians with a FWHM of 0.1~eV of both initial and spawned trajectory basis functions on their respective electronic state. This is done at each time step and the spectrum is obtained by weighting the trajectory basis functions by their population. Lastly, a 50~fs Gaussian time broadening is applied to facilitate comparison to experiment.

All of the above calculations were performed using the TeraChem quantum chemistry package\cite{seritan2020terachem,seritan2021terachem} and the FMS program \cite{levine_implementation_2008}. 

Lastly, we investigate the potential change in x-ray absorption associated with ISC from the S$_1$ state to the T$_1$ state, on which the excited state population might be trapped. To this end, we perform calculations at the OO-DFT/SCAN level of theory\cite{sun2015strongly,hait2021orbital} with scalar relativistic effects included using the spin-free exact two component model\cite{cunha2022relativistic} in QChem\cite{epifanovsky2021software} (using the aug-pcX-2 basis\cite{ambroise2018probing} on O and aug-pcseg-1 basis\cite{jensen2014unifying} on all other atoms), using the respective S$_0$, S$_1$, and T$_1$ BHandHLYP/def2-SVP optimized geometries. We determine the energies of the S$_0$, S$_1$, and T$_1$ states as well as the energies of states corresponding to oxygen 1s core excitation from these states following the procedure outlined in previous studies\cite{hait2021orbital,cunha2022relativistic}. Finally, we calculate transition strengths for the core excitations of the S$_1$ and T$_1$ states to probe differences in spectral intensity obtained from ISC from S$_1$ to T$_1$.

\section{Simulated Excited State Populations}\label{SI:populations}

\begin{figure}[!h]
    \centering
    \includegraphics[width=\textwidth]{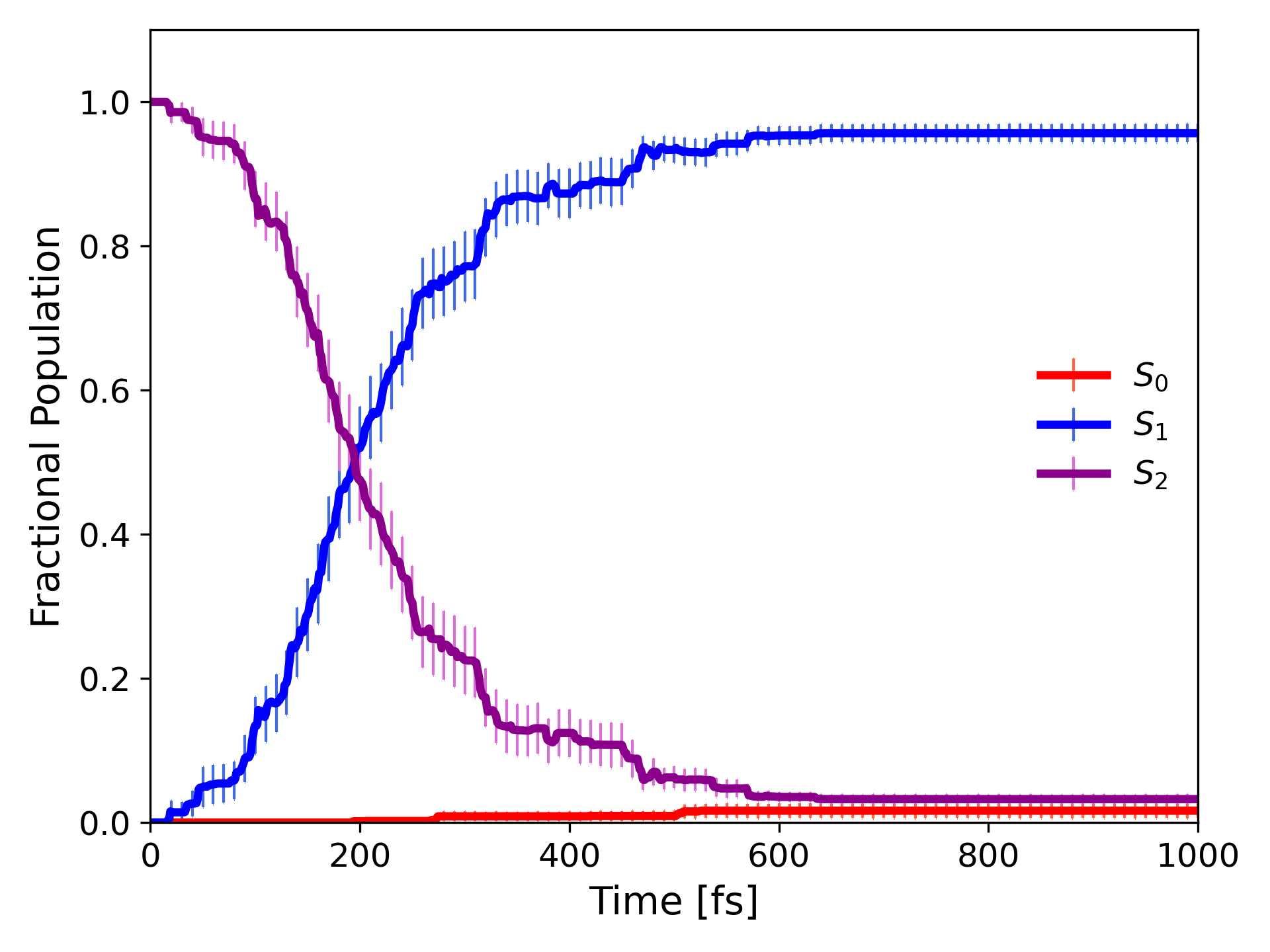}
    \caption{Simulated populations of the S$_0$, S$_1$, and S$_2$ states during the AIMS simulation following initial excitation to the S$_2$ state. The plotted error bars were obtained from Bootstrap sampling.}
    \label{fig:ES_populations}
\end{figure}

\begin{figure}[!h]
    \centering
    \includegraphics[width=\textwidth]{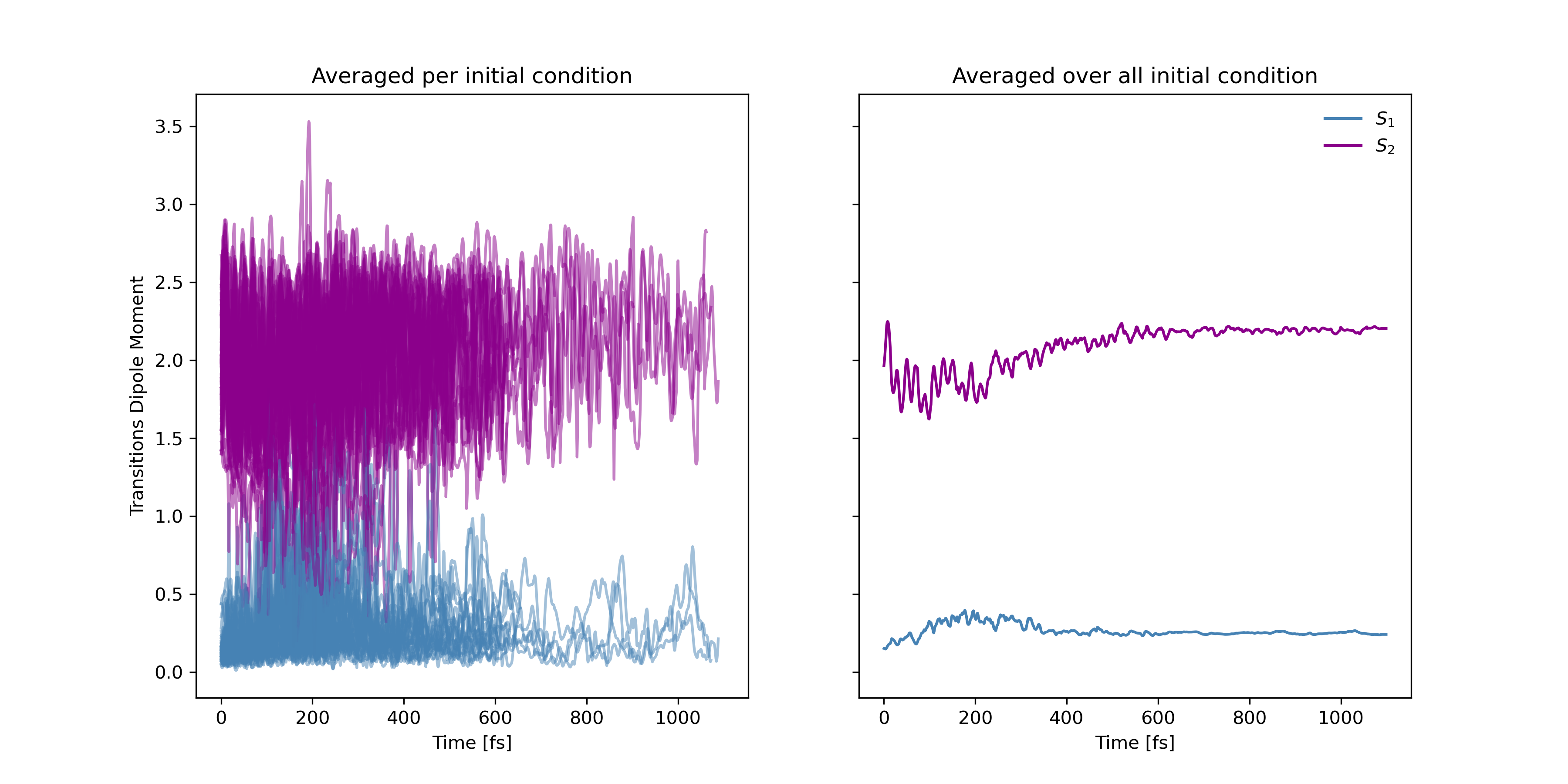}
    \caption{Simulated transition dipole moments for transition from S$_0$ to S$_1$ and S$_2$ states during the AIMS simulation. (Left) Averaged transition dipole moments for each of the 50 initial conditions. (Right) Averaged transition dipole moments over all trajectories.}
    \label{fig:ES_tdms}
\end{figure}

Using AIMS, we are able to track the population of different electronic states included in the nonadiabatic dynamics over time following the initial excitation to the S$_2$ ($\pi\pi^*$) state. In Figure \ref{fig:ES_populations}, we display the transient population over the 1~ps AIMS dynamics with error bars estimated from Bootstrap sampling. As seen the S$_2$ state rapidly depopulates within the first 500~fs by internal conversion to the S$_1$ ($n\pi^*$) state on which most of the population remains for the remainder of the simulation. A very small fraction of the total population does reach the ground state, but this is only the case for 8 out of the 366 resulting trajectory basis functions.
To verify that the mapping of the $n\pi^*$ and $\pi\pi^*$ states to the S$_1$ and S$_2$ states is consistent across the dynamics, we show the average transition dipole moments of these for each initial condition and the total average in Figure \ref{fig:ES_tdms}.
As seen from the averaged transitions dipole moments, the S$_2$ state is generally bright, which is consistent with it being the $\pi\pi^*$ state, while S$_1$ is generally dark consistent with the $n\pi^*$ assignment. This is evident both for individual initial conditions and across all of the trajectories. It is only in the time region where decay from S$_2$ to S$_1$ happens that the two states are close in energy and flip very briefly. Nevertheless, mapping of $n\pi^*$ and $\pi\pi^*$ states to the S$_1$ and S$_2$, respectively, is generally preserved.

\section{Simulated \texorpdfstring{$^1\pi\pi^{*}$}{pi-pi*} state signature} \label{SI:S2state}

\begin{figure}[!h]
    \centering
    \includegraphics[width=\textwidth]{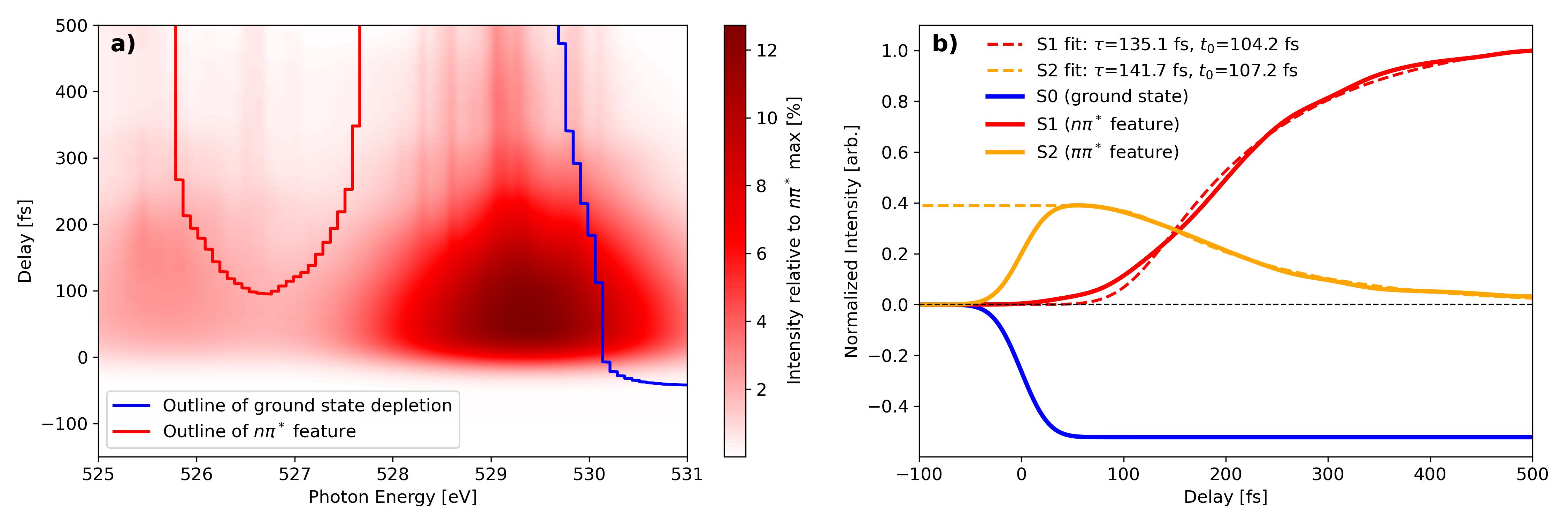}
    \caption{Comparison of excited state signatures. a) Simulated transient NEXAFS spectrum for the initially occupied S$_2$ state with $\pi\pi^{*}$ character. The red and blue outline mark the position of the S$_1$ feature and S$_0$ depletion, respectively. Both outlines are taken form figure 3 c). The intensity is normalized to the maximum of the S$_1$ state with $n\pi^{*}$ character and plotted as relative percentage. The $\pi\pi^{*}$ state has two components, one with an maximum intensity of $\approx$ 13 \% around 529.3~eV, that overlaps with the ground state depletion. And a weak contribution with a maximum intensity of $\approx$ 3 \% at 525.6~eV which corresponds to the direct excitation of a oxygen 1s electron to the molecular orbital with $\pi$ character. b) Integrated lineouts of the individual states. The fits for the S$_1$ and S$_2$ state show that the transition from one state to the other occurs simultaneous at 104.2~fs and 107.2~fs, with comparable time constants of 135.1~fs and 141.7~fs, respectively.}
    \label{fig:S2}
\end{figure}

The AIMS method used to simulate the excited-state dynamics enables analysis of the transient absorption spectra associated with each electronic state independently. The dominant feature in both the experimental and simulated data is the signature of the $n\pi^*$ state, which exhibits a delayed onset. The initially populated $\pi\pi^*$ state is challenging to observe experimentally due to its weak resonant excitation cross-section.
Figure~\ref{fig:S2} a) shows the simulated transient NEXAFS spectrum of acetophenone, considering only the initially excited $\pi\pi^*$ state. The intensity is plotted as a relative percentage with respect to the maximum signal of the more pronounced $n\pi^*$ state. The positions of the $n\pi^*$ and ground-state depletion features (taken from Figure 3 c) in the main text) are indicated by blue and red outlines, respectively. The $\pi\pi^*$ state exhibits two primary contributions: the dominant feature is centered at 529.3~eV with a maximum relative intensity of approximately 13\%, overlapping with the ground-state depletion at 531~eV. This overlap results in a narrowing of the depletion feature during the first 300~fs. A weaker contribution, with a relative intensity of approximately 3\%, is centered at 525.6~eV, lying below the $n\pi^*$ signature at 526.7~ eV.
Electronic structure calculations predict the $n\pi^*$ state to be the lowest in energy, corresponding to an electron displacement from the HOMO to the LUMO. In contrast, the $\pi\pi^*$ state involves a vacancy in a deeper-lying occupied orbital. As a result, resonant excitation from the oxygen 1s orbital to the $\pi\pi^*$ state is expected to occur at lower photon energy than to the $n\pi^*$ state. Accordingly, the peak at 525.6 eV is assigned to direct excitation from the oxygen 1s level into a $\pi$ orbital, whereas the peak at 529.3~eV corresponds to excitation into the $\pi^*$ orbital of the excited state.
Both features begin to decay with the onset of the $n\pi^*$ state and fully vanish after 300~fs, leaving only minor artifacts (vertical stripes) in the data.

\clearpage
\printbibliography